\documentclass[twoside,twocolumn,9pt]{article}
\usepackage{extsizes}
\usepackage[super,sort&compress,comma]{natbib}
\usepackage[version=3]{mhchem}
\usepackage[left=1.5cm, right=1.5cm, top=1.785cm, bottom=2.0cm]{geometry}
\usepackage{balance}
\usepackage{times,mathptmx}
\usepackage{sectsty}
\usepackage{graphicx}
\usepackage{lastpage}
\usepackage[format=plain,justification=raggedright,singlelinecheck=false,
font={stretch=1.125,small,sf},labelfont=bf,labelsep=space]{caption}
\usepackage{float}
\usepackage{fancyhdr}
\usepackage{fnpos}
\usepackage[english]{babel}
\usepackage{array}
\usepackage{droidsans}
\usepackage{charter}
\usepackage[T1]{fontenc}
\usepackage[usenames,dvipsnames]{xcolor}
\usepackage{setspace}
\usepackage[compact]{titlesec}

\usepackage[outdir=./]{epstopdf}

\usepackage{amssymb}
\usepackage{amsmath}
\usepackage{mathtools} 
\usepackage{bm}
\usepackage{wasysym}
\usepackage{enumitem}
\usepackage{hyperref} 

\newcommand{\lla}{\left\langle}
\newcommand{\rra}{\right\rangle}

\newcommand{\brh}{\mathrm{\bf H}}

\newcommand{\vphi}{\varphi}
\newcommand{\ttau}{\tilde \tau}
\newcommand{\ttone}{\tilde \tau_{1}}
\newcommand{\lbar}[1]{\overline{#1}}
\newcommand{\dst}{\displaystyle}

\graphicspath{{./figures/},{./Images/}}

\newcommand{\ocite}[1]{\hspace{-1 ex} \nocite{#1}\citenum{#1}}

\newcommand{\onlinecite}[1]{\hspace{-1 ex} \nocite{#1}\citenum{#1}}

\graphicspath{{figures/}}

\definecolor{cream}{RGB}{222,217,201}

\begin{document}

\pagestyle{fancy}
\thispagestyle{plain}
\fancypagestyle{plain}{

}

\makeFNbottom
\makeatletter
\renewcommand\LARGE{\@setfontsize\LARGE{15pt}{17}}
\renewcommand\Large{\@setfontsize\Large{12pt}{14}}
\renewcommand\large{\@setfontsize\large{10pt}{12}}
\renewcommand\footnotesize{\@setfontsize\footnotesize{7pt}{10}}
\makeatother

\renewcommand{\thefootnote}{\fnsymbol{footnote}}
\renewcommand\footnoterule{\vspace*{1pt}%
\color{cream}\hrule width 3.5in height 0.4pt \color{black}\vspace*{5pt}}
\setcounter{secnumdepth}{5}

\makeatletter
\renewcommand\@biblabel[1]{#1}
\renewcommand\@makefntext[1]%
{\noindent\makebox[0pt][r]{\@thefnmark\,}#1}
\makeatother
\renewcommand{\figurename}{\small{Fig.}~}
\sectionfont{\sffamily\Large}
\subsectionfont{\normalsize}
\subsubsectionfont{\bf}
\setstretch{1.125} 
\setlength{\skip\footins}{0.8cm}
\setlength{\footnotesep}{0.25cm}
\setlength{\jot}{10pt}
\titlespacing*{\section}{0pt}{4pt}{4pt}
\titlespacing*{\subsection}{0pt}{15pt}{1pt}

\fancyfoot{}
\fancyfoot[LO,RE]{\vspace{-7.1pt}\includegraphics[height=9pt]{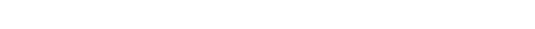}}
\fancyfoot[CO]{\vspace{-7.1pt}\hspace{13.2cm}\includegraphics{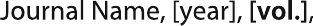}}
\fancyfoot[CE]{\vspace{-7.2pt}\hspace{-14.2cm}\includegraphics{head_foot/RF}}
\fancyfoot[RO]{\footnotesize{\sffamily{1--\pageref{LastPage} ~\textbar  \hspace{2pt}\thepage}}}
\fancyfoot[LE]{\footnotesize{\sffamily{\thepage~\textbar\hspace{3.45cm} 1--\pageref{LastPage}}}}
\fancyhead{}
\renewcommand{\headrulewidth}{0pt}
\renewcommand{\footrulewidth}{0pt}
\setlength{\arrayrulewidth}{1pt}
\setlength{\columnsep}{6.5mm}
\setlength\bibsep{1pt}

\makeatletter
\newlength{\figrulesep}
\setlength{\figrulesep}{0.5\textfloatsep}

\newcommand{\topfigrule}{\vspace*{-1pt}%
\noindent{\color{cream}\rule[-\figrulesep]{\columnwidth}{1.5pt}} }

\newcommand{\botfigrule}{\vspace*{-2pt}%
\noindent{\color{cream}\rule[\figrulesep]{\columnwidth}{1.5pt}} }

\newcommand{\dblfigrule}{\vspace*{-1pt}%
\noindent{\color{cream}\rule[-\figrulesep]{\textwidth}{1.5pt}} }

\makeatother

\twocolumn[
  \begin{@twocolumnfalse}
\vspace{3cm}
\sffamily
\begin{tabular}{m{4.5cm} p{13.5cm} }

\includegraphics{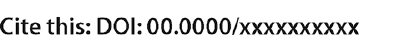} & \noindent\LARGE{\textbf{Active Brownian Filaments with Hydrodynamic Interactions: Conformations and Dynamics}} \\
\vspace{0.3cm} & \vspace{0.3cm} \\

 & \noindent\large{ Aitor Mart\'{i}n-G\'{o}mez, Thomas Eisenstecken, Gerhard Gompper, and Roland G. Winkler} \\

\includegraphics{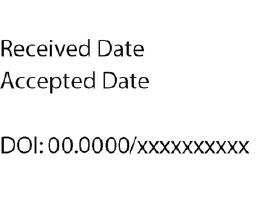} & \noindent\normalsize

{The conformational and dynamical properties of active self-propelled filaments/polymers  are investigated in the presence of hydrodynamic interactions by both, Brownian dynamics simulations and analytical theory. Numerically, a discrete linear chain composed of active Brownian particles is considered, analytically, a continuous linear semiflexible polymer with active velocities changing diffusively. The force-free nature of active monomers is accounted for---no Stokeslet fluid flow induced by active forces---and higher order hydrodynamic multipole moments are neglected. Hence,  fluid-mediated interactions are assumed to arise solely due to intramolecular forces. The hydrodynamic interactions (HI) are taken into account analytically by the preaveraged Oseen tensor, and numerically by the Rotne-Prager-Yamakawa tensor. The nonequilibrium character of the active process implies a dependence of the stationary-state properties on HI via the polymer relaxation times. In particular, at moderate activities, HI lead to a substantial shrinkage of flexible and semiflexible polymers to an extent far beyond shrinkage of comparable free-draining polymers; even flexible HI-polymers shrink, while active free-draining polymers swell monotonically. Large activities imply a reswelling, however, to a less extent than for non-HI polymers, caused by the shorter polymer relaxation times due to hydrodynamic interactions. The polymer mean square displacement is enhanced, and an activity-determined ballistic regime appears. Over a wide range of time scales, flexible active polymers exhibit a hydrodynamically governed subdiffusive regime, with an exponent significantly smaller than that of the Rouse and Zimm models of passive polymers. Compared to simulations, the approximate analytical approach predicts a weaker hydrodynamic effect. Overall, hydrodynamic interactions modify the conformational and dynamical properties of active polymers substantially.

}  \\%

\end{tabular}

\end{@twocolumnfalse} \vspace{0.6cm}

  ]

\renewcommand*\rmdefault{bch}\normalfont\upshape
\rmfamily
\section*{}
\vspace{-1cm}

\
\footnotetext{Theoretical Soft Matter and Biophysics, Institute for
Advanced Simulation and Institute of Complex Systems,
Forschungszentrum J\"ulich, D-52425 J\"ulich, Germany; E-mail: r.winkler@fz-juelich.de; g.gompper@fz-juelich.de}


\section{Introduction} \label{sec:introduction}

The perpetual conversion of either internal chemical energy, or utilization of energy from the environment, into directed motion is a key feature of active matter \cite{elge:15,bech:16}. Its respective out-of-equilibrium nature  is the origin of intriguing emerging structural and dynamical properties, which are absent in passive systems. This particularly applies to soft matter systems, e.g., comprised of filaments or polymers, which renders active soft matter a promising  class of new materials \cite{cate:11,need:17}. Nature provides various examples of filamentous, polymer-like active agents or phenomena where activity governs the nonequilibrium dynamics of passive molecules. Propelled biological polar semiflexible filaments are ubiquitous, e.g., filamentous actin or microtubules in the cell cytoskeleton due to tread-milling and motor proteins \cite{ridl:03,gang:12}.
In motility assays, filaments are propelled on carpets of motor proteins anchored on a substrate, which results in a directed motion and the appearance of self-organized dynamical patters. \cite{hara:87,nedl:97,scha:10,juel:07,marc:13,pros:15,cord:14,sumi:12,doos:17,wink:17}
A characteristic feature of biological cells is the intrinsic mixture of active and passive components; specifically the active cytoskeleton and a large variety of passive colloidal and polymeric objects. Here, activity implies an  enhanced  random motion of tracer particles \cite{bran:08}.  Furthermore, the active dynamics of microtubules \cite{bran:09} or actin-filaments \cite{webe:15} leads to an accelerated  motion of chromosomal loci \cite{webe:12,jave:13} and chromatin \cite{zido:13}. In addition, ATP-dependent enzymatic activity-induced mechanical fluctuations drive molecular motion in the bacterial cytoplasm and the nucleus of eukaryotic cells \cite{webe:12}. Self-propelled rodlike or semiflexible polymer-like objects are formed via self-assembly, e.g., by dinoflagellates \cite{sela:11,sohn:11}, or grow in bacterial biofilms, such as  {\em Proteus mirabilis} \cite{cope:09}.
Synthetic active or activated colloidal polymers \cite{loew:18} are nowadays synthesized in various ways.
Assembly of active chains of metal-dielectric Janus colloids (monomers) can be achieved by imbalanced interactions, where simultaneously the motility and the colloid interactions are controlled by an AC electric field \cite{yan:16,dile:16,nish:18}. Electrohydrodynamic convection  rolls lead to self-assembled colloidal chains in a nematic liquid crystal matrix and directed movement \cite{sasa:14}. Moreover, chains of linked colloids, which are uniformly coated with catalytic nanoparticles, have been synthesizes \cite{bisw:17}. Hydrogen peroxide decomposition on the surfaces of the colloidal monomers generates phoretic flows, and active hydrodynamic interactions between monomers results in an enhanced diffusive motion \cite{bisw:17}.

Valuable insight into the properties of self-propelled filaments and polymers, or their passive counterparts  embedded in an active environment, is obtained by computer simulations and analytical theory.  Thereby, typically active Brownian polymers (ABPOs), neglecting hydrodynamic interactions (HI) (in the following, we will denote such polymers as ABPOs-HI), have been considered \cite{loi:11,kais:14,hard:14,ghos:14,chel:14,sark:14,shin:15,kais:15,isel:15,sama:16,eise:16,osma:17,eise:17,mart:18.1}, but also particular aspects of fluid-mediate interactions  have been studied \cite{jaya:12,jian:14,jian:14.1,lask:15,pand:16,bisw:17}. Filaments are modeled as semiflexible polymers, with an implementation of activity adapted to the particular propulsion mechanism. Polar polymers, representing actin filaments or microtubules driven by molecular motors, are typically propelled by forces tangential to the polymer contour \cite{chel:14,jian:14,jian:14.1,isel:15,duma:18,prat:18,bian:18}. Here, a sufficiently high activity leads to shrinkage and compactification \cite{isel:15,anan:18}. ABPOs, where every monomer experiences an independent active force whose orientation changes in a diffusive manner \cite{kais:14,eise:16,wink:16}, or passive polymers embedded in an environment of active Brownian particles (ABPs), \cite{hard:14,shin:15} exhibit a different behavior. Flexible ABPOs-HI swell with increasing activity due to local active forces overpowering thermal noise \cite{kais:14,shin:15,eise:16,osma:17,eise:17,mart:18.1}. Semiflexible ABPOs-HI shrink first at moderate activities owing to active intramolecular stresses competing with bending forces, and swell for higher activities similar to flexible ABPOs-HI \cite{hard:14,eise:16,eise:17}. In all cases, a faster dynamics is obtained \cite{isel:15,ghos:14,kais:14,shin:15,eise:16,wink:16,osma:17,eise:17}.

Hydrodynamics changes the properties of active systems in various ways. Since an individual self-propelled particle---an isolated monomer in the case of a colloidal-type polymer \cite{loew:18}---is force and torque free, it creates a flow field lacking a Stokeslet, but includes higher multipole contributions \cite{goet:10,llop:10,yeom:14,lask:15,thee:16.1,zoet:16,wink:17,wink:18}.  Conformational changes and the interference of the monomer flow fields lead to autonomous filament/polymer motion even when individual monomers are non-motile \cite{bisw:17,jaya:12,lask:15}. The conformational and dynamical properties of  polar (actively) driven filaments, which are not force free, are also strongly affected by hydrodynamic interactions \cite{jian:14,jian:14.1}. In particular, hydrodynamic coupling between two filaments leads to cooperative effects \cite{jian:14.1}.

\begin{figure}[t]
\begin{center}
\includegraphics[width=\columnwidth]{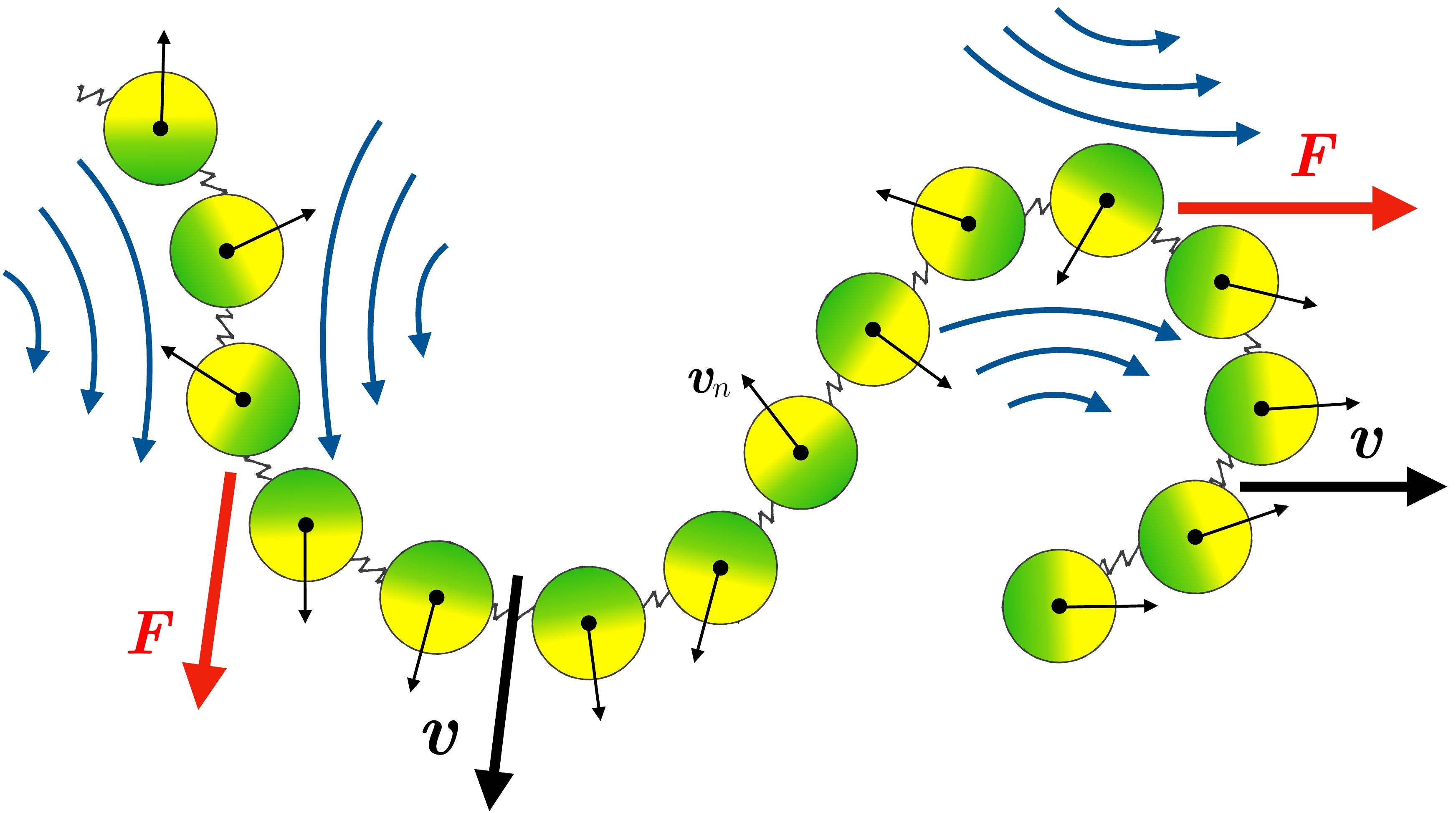}
\end{center}
\caption{Illustration of the activity-induced flow by the motion of an ABPO+HI. Several ABPs moving by chance together in a certain direction, indicated by the velocity arrow $\bm v$, drag other connected ABPs, which in turn  exert a force, $\bm F$, indicated by the red arrow,  on the fluid inducing Stokes flow.  The small arrows $\bm v_n$ display the direction of the active velocity. No flow field is generated by the active motion of an individual ABP. An animation of the dynamics of a discrete ABPO+HI is provided in the ESI.} \label{fig:sketch}
\end{figure}

In this article, we analyze the influence of hydrodynamic interactions on the conformational and dynamical properties of ABPOs, denoted as ABPOs+HI in the following, by computer simulations and an analytical approach.  In simulations, we employ a bead-spring linear phantom or self-avoiding polymer with ABP monomers (cf. Fig.~\ref{fig:sketch}), where the ABP propulsion direction changes diffusively, \cite{eise:17.1} and hydrodynamic interactions are taken into account Rotne-Prager-Yamakawa hydrodynamic tensor. \cite{rotn:69,yama:70}  For the analytical calculations, we consider a Gaussian semiflexible polymer, \cite{wink:94,harn:96,eise:16} with active sites modeled by an Ornstein-Uhlenbeck process (active Ornstein-Uhlenbeck particle, AOUP), \cite{fodo:16,das:18.1,eise:16} where the active velocity vector changes in a diffusive manner; here, HI is included via the preaveragred Oseen tensor. \cite{doi:86,dhon:96} The main purpose of our study is to resolve the influence of hydrodynamics on the properties of self-propelled polymers, respecting  the force-free nature of an individual active agent. Hence, no Stokeslet due to self-propulsion is present. Only Stokeslets arising from bond, bending, and excluded-volume interactions between monomers, as well as thermal forces are considered. Moreover,  we neglect higher order multipole contributions of the active monomers, especially the force dipole. Since we consider point particles, source multipoles are also absent. All these multipoles decay faster than a Stokeslet. Hence, we capture the long-range character of HI in polymers of a broad class of active monomers. As far as near-field hydrodynamic effects are concerned, our model closest resembles a polymer composed of neutral squirmers \cite{llop:10,goet:10,thee:16,wink:16.1}, where particular effects by higher multipole interactions between monomers are not resolved.  \cite{jaya:12,lask:15}

Our studies reveal a decisive influence of hydrodynamic interactions on the polymer conformations and dynamics. In particular, even flexible ABPOs+HI shrink at moderate activities, where ABPOs-HI swell monotonically. At high activities, ABPOs+HI swell, but to an extent, which is considerably smaller than that of ABPOs-HI. This indicates a dependence of the stationary-state distribution function on hydrodynamics, an effect absent for passive systems. The reason is the violation of the fluctuation-dissipation theorem of the active processes, which leads to the dependence of stationary-state properties on the hydrodynamically modified relaxation times. The shrinkage is then a consequence of the time-scale separation between the thermal process, dominating for zero or very weak activities, and the active process with hydrodynamically accelerated relaxation times. The modified, activity-dependent relaxation times also affect the translation motion, and a subdiffusive time regime appears, where the mean square displacement (MSD) exhibits a power-law dependence with the exponent $\alpha'= 2/5$, significant smaller  than the Zimm value, $\alpha' =2/3$, of a passive polymer.

The manuscript is organized as follows. Section~\ref{sec:discrete} describes the discrete model of the ABPO along with the simulation approach, and presents simulation results. Section~\ref{sec:model} describes the continuum model of an active polymer, its analytical solution, and discusses  conformational and dynamial properties.  Section~\ref{sec:summary} summarizes our findings.

\section{Computer simulations} \label{sec:discrete}

\subsection{Model}

A semiflexible active polymer is composed  of $N_m$ active Brownian particles (ABPs) ($i=1,\ldots,N_m$, cf. Fig.~\ref{fig:sketch}),\cite{kais:14,eise:17.1} which obey the  equations of motion
\begin{align} \label{eq:langevin_bd}
  \dot{\bm{r} }_i(t) = & \ v_0 \bm e_i(t)+ \sum^{N_m}_{j=1} \mathbf{H}_{ij} \left[ \bm{ F}_i(t) + \bm {\varGamma}_i(t) \right]  , \\ \label{eq:orientation}
  \dot{\bm{e} }_i(t)= & \ \hat{\bm{\eta}}_i(t)\times  \bm{e}_i(t).
\end{align}
Here, $\bm r_i(t)$ and $\dot {\bm r}_i(t)$ denote the position and velocity of particle $i$, respectively,  and $\bm{ v } _i(t) = v_0 \bm{ e } _i(t)$ is the active velocity with the propulsion direction $\bm e_i$ ($|\bm e_i|=1$), which changes in a diffusive manner according to Eq.~(\ref{eq:orientation}).   The forces $\bm{F}_i(t)= - \bm{\nabla}_{\bm{ r } _i} (U_l + U_b + U_{LJ})$ following from the bond ($U_l$), bending ($U_b$), and volume exclusion ($U_{LJ}$) potentials, \cite{eise:17.1}
\begin{align} \label{eq:pot_bond}
U_{l} = & \ \frac{\kappa_{l}}{2} \sum^{N_m}_{i=2} \left(|\bm{R}_i|  - \ l \right)^2 , \\ \label{eq:pot_bend}
U_{b} = & \ \frac{\kappa_{b}}{2} \sum^{N_m -1}_{i=2} \left(   \bm{R}_{i+1} -  \bm{R}_{i}  \right)^2 , \\ \label{eq:pot_lj}
U_{LJ} = & \
\left\{ \begin{matrix} 4 \epsilon \dst \sum_{i<j} \left[
\left( \dst \frac{\dst \sigma}{\dst  r_{ij}} \right)^{12} - \left( \dst \frac{\dst \sigma}{\dst r_{ij}} \right)^6 + \dst \frac{1}{4}\right],&r_{ij} < \sqrt[6]{2}\sigma \\
0, &r_{ij} > \sqrt[6]{2} \sigma
\end{matrix}
\right.  ,
\end{align}
where $\bm R_{i+1} = \bm r_{i+1} - \bm r_i$ is the bond vector,  $\bm{r}_{ij}=\bm{r}_{i}-\bm{r}_{j}$ the vector between monomers $i$ and $j$, and $r_{ij} = |\bm r_{ij}|$. The energy $\epsilon$ measures the stength of the purely repulsive potential, and $\sigma$ is the diameter of a monomer.
$\bm \varGamma_i$ and $\hat{\bm \eta}_i$ are Gaussian and Markovian stochastic processes with zero mean and the second moments
\begin{align}
\lla \bm \varGamma_i (t) \bm \varGamma_j^T (t')\rra & = 2 k_BT \mathbf{H}_{ij}^{-1} \delta(t-t') \ , \\
\lla \hat \eta_{i \alpha}(t) \hat \eta_{j \beta}(t') \rra & = 2 D_R \delta_{\alpha \beta} \delta_{ij} \delta(t-t')) \ ,
\end{align}
where $\bm \varGamma_i^T$ denotes the transpose of $\bm \varGamma_i$, and $\mathbf{H}_{ij}^{-1}$ the inverse of $\mathbf{H}_{ij}$; $T$ is the temperature, $k_B$ the Boltzmann constant, and $D_R$ the rotational diffusion coefficient of a spherical colloid.   The tensor $\mathbf{H}_{ij}(\bm r_{ij}) = \delta_{ij} \mathbf{I}/3 \pi \eta l + (1-\delta_{ij}) \bm{\Omega}(\bm r_{ij})$ accounts for hydrodynamic interactions, with the first term including local friction, and the Rotne-Prager-Yamakawa tensor \cite{rotn:69,yama:70,jain:12.1}
\begin{align}
\bm{\Omega}\left( \bm{r}_{ij} \right ) = \left\{
\begin{matrix}
\dst \frac{1}{8\pi\eta r_{ij}} \left[ \mathbf{I} + \frac{\bm{r}_{ij} \bm{r}_{ij}^T}{r^2_{ij}} + \frac{l^2}{2r^2_{ij}} \left( \frac{1}{3} \mathbf{I} - \frac{\bm{r}_{ij}  \bm{r}_{ij}^T }{r^2_{ij}}  \right) \right] ,\ & r_{ij} > l \\[2ex]
\dst \frac{1}{3 \pi\eta l} \left[  \left ( 1 - \frac{9}{16} \frac{r_{ij}}{l} \right ) \mathbf{I}  + \frac{3}{16} \frac{r_{ij}}{l} \frac{\bm{r}_{ij} \bm{r}_{ij}^T}{r^2_{ij}}  \right]  \ , & \ r_{ij} < l
\end{matrix}
\right. ,
\end{align}
with the solvent viscosity $\eta$  and  the unit matrix $\bf I$.
We assume a touching bead model of spherical colloids, hence, the monomer hydrodynamic radius is half of the bond length $l$. The Rotne-Prager-Yamakawa tensor insures the positive definiteness of the hydrodynamic tensor even at small distances.

The translational equations of motion \eqref{eq:langevin_bd} are solved via the Ermark-McCammon algorithm. 	\citep{alle:87,erma:78} The procedure to solve the equations of motion (\ref{eq:orientation}) for the orientation vectors is described in Sec.~S-IV of the ESI. \cite{wink:15}

We characterize activity by the P\'eclet number $Pe$ and the ratio $\Delta$ between translational, $D_T=k_BT/3\pi \eta l$,  and rotational, $D_R$, diffusion coefficient of an isolated monomer, where
\begin{align} \label{eq:peclet}
Pe = \frac{v_0}{l D_R} , \hspace*{5mm} \Delta = \frac{D_T}{l^2 D_R} .
\end{align}
The coefficient  $\kappa_{l}$ (Eq.~\eqref{eq:pot_bond}) for the bond strength is adjusted according to the applied P\'eclet number, in order to avoid bond stretching with increasing activity. By choosing $\kappa_{l} l^2 /k_BT = (10  +  2 Pe ) 10^3$,  bond-length variations are smaller than $3 \%$ of the equilibrium value $l$. Furthermore, the scaled bending force coefficient $\tilde{\kappa}_{b}=\kappa_{b} l^2 /k_BT$  (Eq.~\eqref{eq:pot_bend}) is related to the polymer persistence length, $l_p=1/(2p)$, by
\begin{equation}
pL = N_m \frac{\tilde{\kappa}_{b} \left( 1 - \coth \left( \tilde{\kappa}_{b} \right) \right) +1}{
\tilde{\kappa}_{b} \left( 1 + \coth \left( \tilde{\kappa}_{b} \right) \right) -1} \ \  .
\end{equation}
The parameters of the truncated and shifted Lennard-Jones potential are $\sigma = 0.8 l$ and $\epsilon = k_BT$.

\subsection{Conformations} \label{sec:sim_conformations}

\begin{figure}[t!]
\begin{center}
\includegraphics[width=0.9\columnwidth]{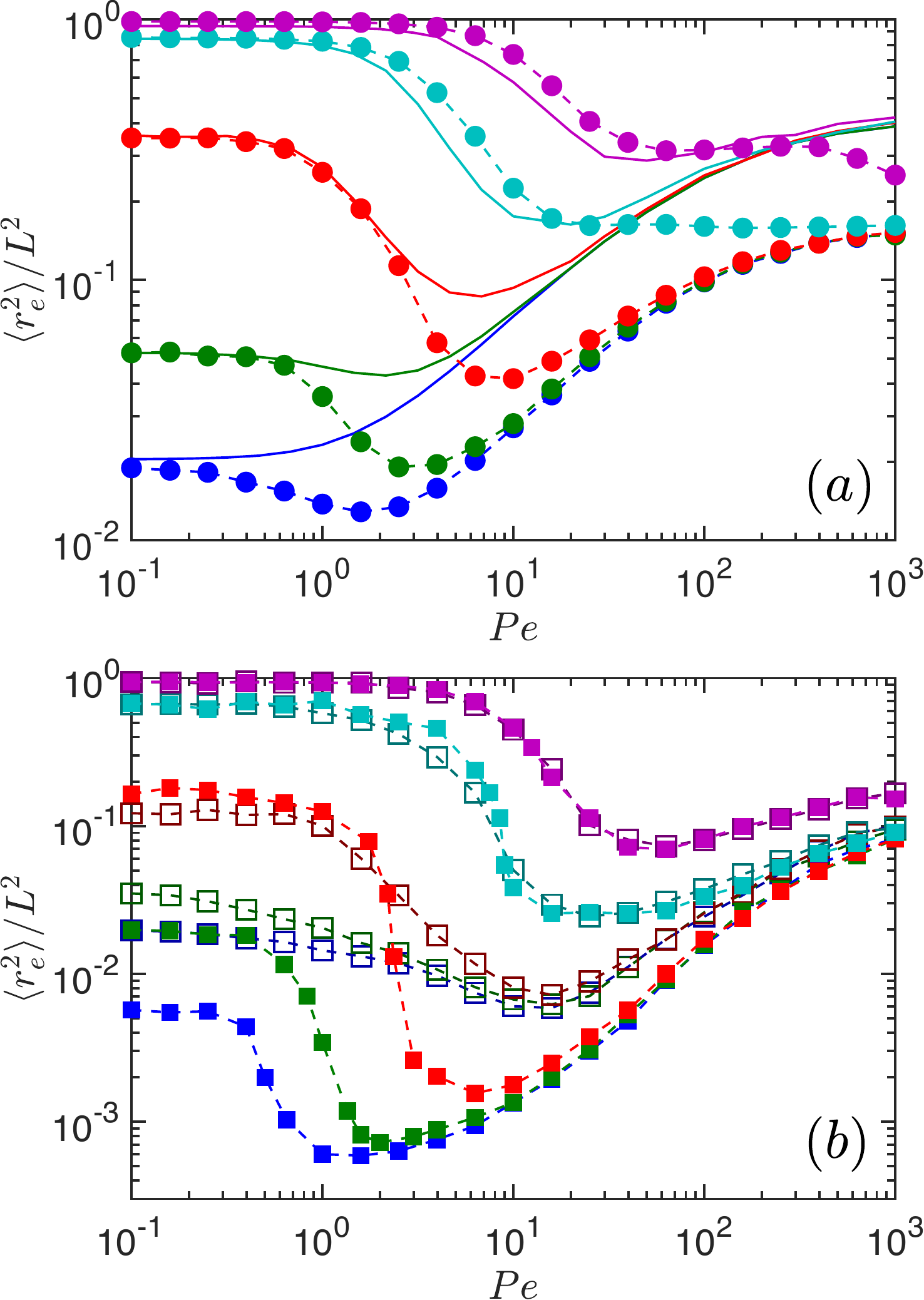}
\end{center}
\caption{Mean square end-to-end distance (simulations) as a function of the P\'eclet number for semiflexible polymers  with (a) $N_m=50$ ($L=49l$)  monomers (bullets) and $pL = 5 \times 10^1$ (blue), $1.5 \times 10^1$ (green), $2.6$ (red), $2.5 \times 10^{-1}$ (cyan), and $2.5 \times 10^{-2}$ (purple), and (b) $N_m=200$ ($L=199l$)  monomers (squares)  for
$pL = 2 \times 10^2$ (blue), $6 \times 10^1$ (green), $10^1$ (red), $1$ (cyan), and $10^{-1}$ (purple). In (a), the solid lines are theoretical results for ABPOs-HI, and bullets are for phantom polymers. In (b) filled squares correspond to phantom and open squares to  self-avoiding polymers. The dashed lines are guides for the eye.
See Fig.~\ref{fig:snapshot} for snapshots and the ESI for a movie.} \label{fig:end-sim+HI}
\end{figure}

We characterize the polymer conformations by the mean square end-to-end distance. Results for phantom polymers of length $L=(N_m-1)l=49l$ and $L=199l$ are presented in Fig.~\ref{fig:end-sim+HI}. Evidently, ABPOs in the presence of hydrodynamic interactions exhibit a pronounced shrinkage for $1 \lesssim Pe \lesssim 10$, where shrinkage depends on polymer length and is substantially stronger for longer polymers.
Semiflexible ABPOs+HI shrink stronger than ABPOs-HI, but the effect vanishes gradually as $pL \to 0$. This is a consequence of the reduced influence of hydrodynamic interactions for rather stiff polymers. \cite{wink:07.1} Yet, the asymptotic swollen value for $Pe \to \infty$ of ABPOs+HI is smaller than the value for ABPOs-HI, for which theory predicts $L^2/2$ and simulations yield approximately $2L^2/5$. \cite{eise:16}  Hence, hydrodynamic interactions affect the swelling behavior of flexible and semiflexible polymers for all $Pe >0$. In particular,  the asymptotic size $\langle \bm r_e^2 \rangle \approx L^2/10$  for $Pe \to \infty$, which is independent of stiffness, is smaller than the value for an ABPO-HI.

\begin{figure}[t!]
\begin{center}
\includegraphics[width=\columnwidth]{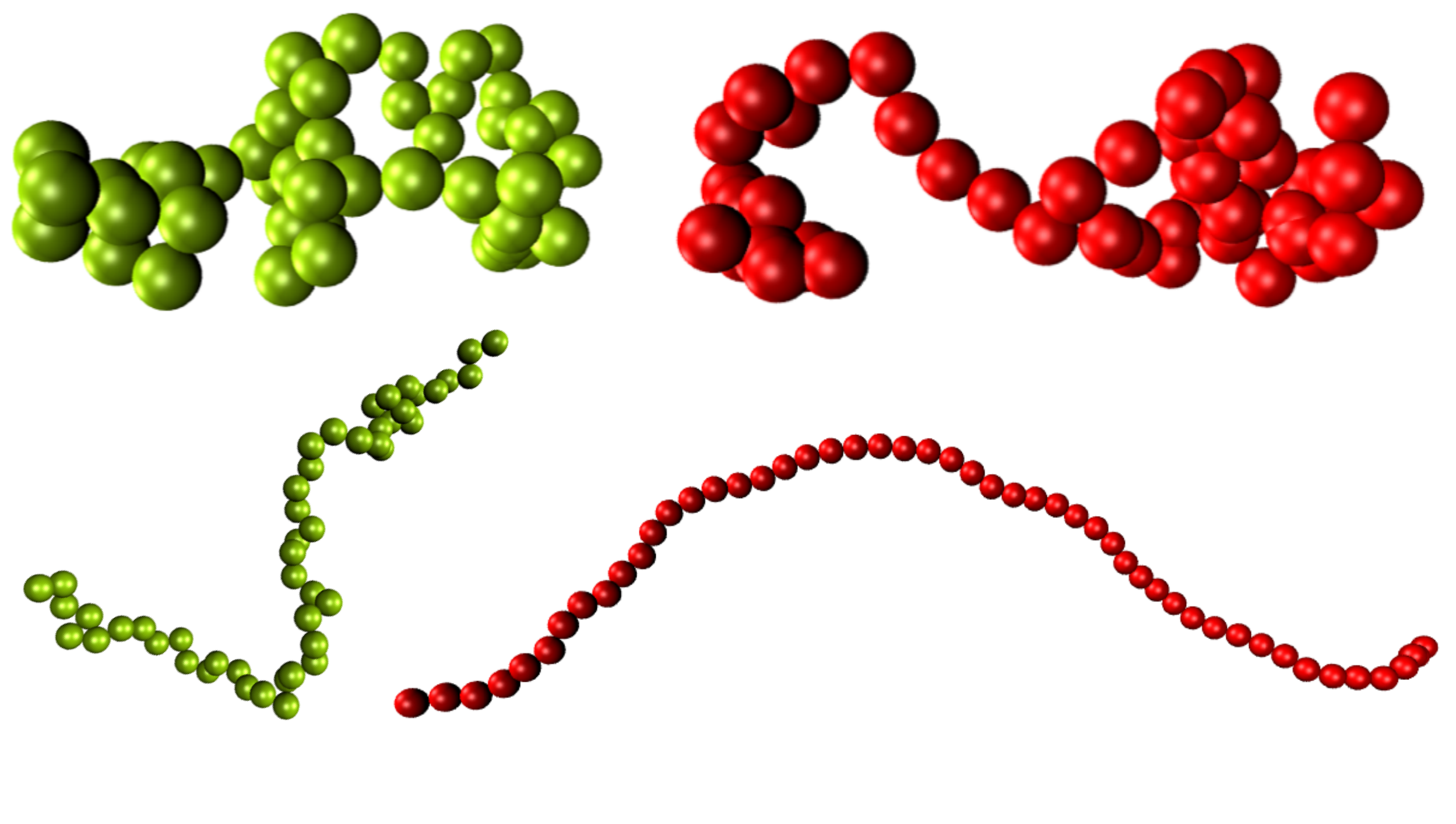}
\end{center}
\caption{Configurations of flexible phantom ABPOs of length $N_m=50$ for the P\'eclet numbers $Pe=1$ (top) and $10^3$ (bottom) in the presence  (green) and absence (red) of hydrodynamic interactions. A movie of an ABPO+HI is available at the ESI.} \label{fig:snapshot}
\end{figure}

Self-avoidance reduces the extent of shrinkage, specifically of flexible polymers. This is illustrated in Fig.~\ref{fig:end-sim+HI}(b). For $pL\gtrsim 10$, the equilibrium value $\langle \bm r_e^2 \rangle$ of a self-avoiding polymer is swollen compared to a phantom polymer. Such an ABPO+HI exhibits a less pronounced shrinkage for all polymer lengths. Naturally, excluded-volume effects vanish with decreasing $pL$, and for $pL<1$ there is hardly any difference between a phantom and a self-avoiding polymer. Moreover, the swelling behavior with and without excluded-volume interactions is rather similar in the limit $Pe \gg 1$. Interestingly, phantom and self-avoiding polymers show a universal dependence on $Pe$ as they start to swell. Here, active forces exceed both, excluded-volume interactions and bending forces. As predicted by theory (cf. Sec. \ref{sec:relax_time}), the internal dynamics is determined by the modes of a flexible polymer, i.e., intermolecular tension, in this regime. Snapshots of conformations with and without HI are presented in Fig.~\ref{fig:snapshot}.

\subsection{Dynamical Properties}
\begin{figure}[t!]
\begin{center}
\includegraphics[width=\columnwidth]{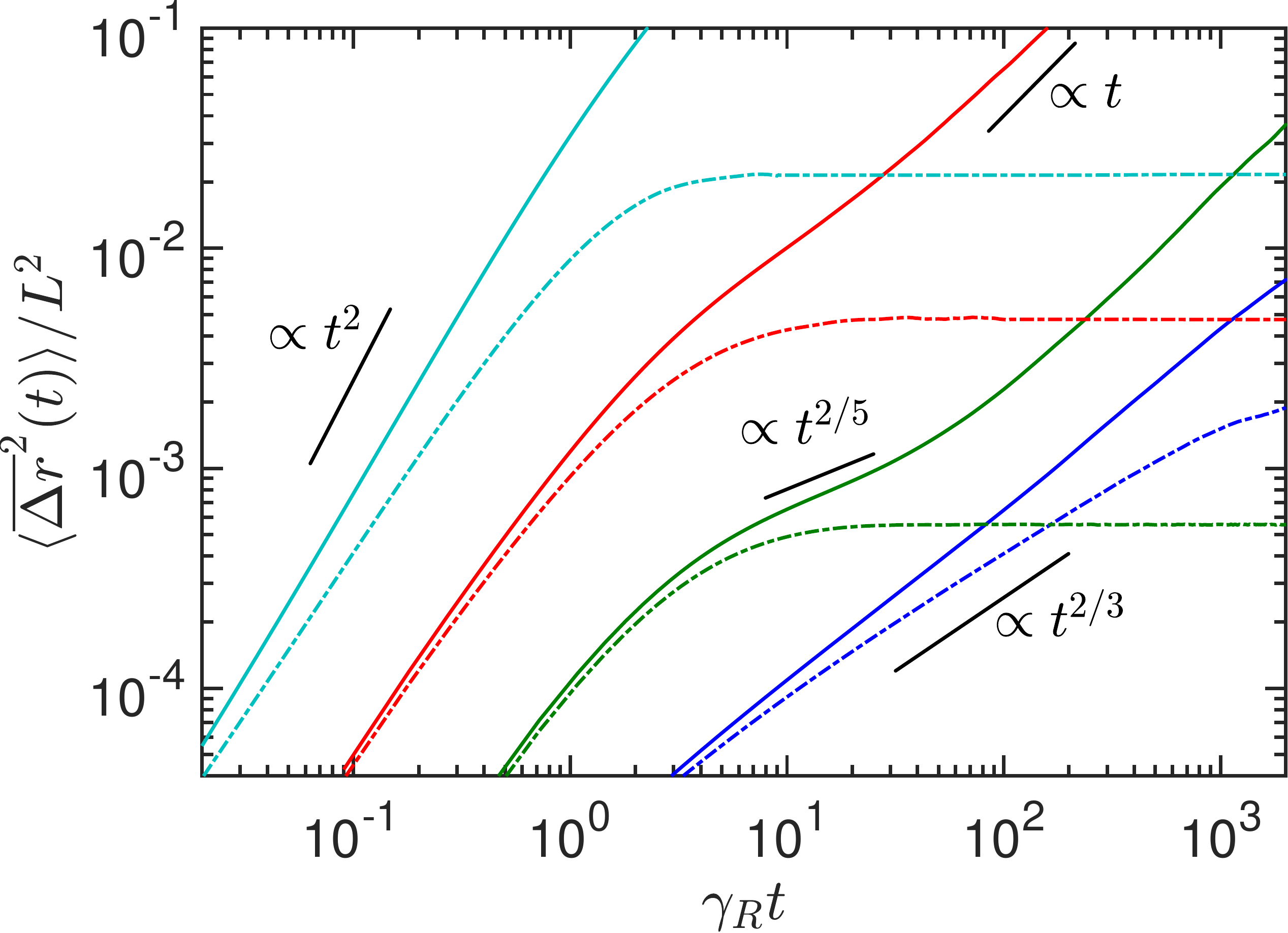}
\end{center}
\caption{Mean square displacement of a flexible phantom polymer with  $N_m=200$ ($pL=200$) monomers for the P\'eclet numbers  $Pe = 0$ (blue), $10^1$ (green), $10^2$ (red), and $10^3$ (cyan). The time is scaled by the factor $\gamma_R = 2 D_R$. The solid lines indicate the monomer MSD and the dashed lines the MSD in the polymer center-of-mass reference frame. The black lines are guides for the eye correspond to a power-law fit of the data in the respective regime.
} \label{fig:msd-HI}
\end{figure}

The dynamics of ABPO+HI is characterized by the monomer mean square displacement (MSD) averaged over all monomers   $\langle \lbar{\Delta \bm r^2}(t) \rangle  = \sum_{i} \langle (\bm r_i(t) - \bm r_i(0))^2 \rangle /N_m$.
Figure~\ref{fig:msd-HI} shows MSDs of a polymer with $N_m=200$ monomers for various P\'eclet numbers. A passive polymer exhibits the well-known Zimm behavior, with the time dependence $t^{2/3}$ of the MSD in the center-of-mass reference frame for $t/\tau_Z \ll 1$.  At long times $t/\ttone \gg 1$,  the center-of-mass displacement dominates the monomer MSD for all P\'eclet numbers. Here, we find the HI-independent MSD $\langle \bm r^2_{cm} \rangle = 2 v_0^2 l t/ \gamma_R L$ following from Eq.~\eqref{eq:langevin_bd} for $Pe \gg 1$ (see also Eq.~\eqref{eq:amp_CMmsd_SP}).
For P\'eclet numbers $Pe>1$, the active ballistic regime, $\langle \lbar{\Delta  \bm r^2} \rangle \sim t^2$, is present at short times $(\gamma_Rt, t/\tilde \tau_1 < 1)$. Moreover,  for $t/\ttone \gtrsim 1$ and moderate P\'eclet numbers, $Pe \approx 10$, activity implies a polymer-specific regime, where the monomer MSD exhibits a power-law dependence $\langle \lbar{\Delta \bm r^2}(t) \rangle  \sim t^{\alpha'}$,  with an exponent of $\alpha' \approx 2/5$, a value smaller than the exponent $\alpha' =2/3$ of the Zimm dynamics. The reduction of the exponent is a clear consequence of the coupling between hydrodynamics and activity, since ABPOs-HI always display slopes $\alpha' \gtrsim 1/2$, where $\alpha' = 1/2$ is the value of a passive flexible polymer (Rouse model). \cite{eise:17} However, this regime appears as a crossover from the ballistic to the diffusive regime. Nevertheless, it is a consequence of hydrodynamics with a sub-diffusive motion. The polymer-specific regime vanishes gradually with increasing $Pe$. As discussed in Sec.~\ref{sec:relax_time}, this is a consequence of the decreasing polymer  relaxation times with increasing activity.

\section{Analytical Model of Active Brownian Polymer in Solution} \label{sec:model}

The simulations of the previous section yield a surprising shrinkage of even flexible polymers by hydrodynamic interactions. In order to shed light on the underlying mechanisms, we study an mean-field analytical model, where
an active polymer is described by a continuous  Gaussian semiflexible polymer model. \cite{wink:94,harn:96} This approach has been applied successfully in the analysis the properties of ABPO-HI \cite{eise:16,eise:17} in close quantitative agreement with simulations.\cite{eise:17.1}

\subsection{Equations of Motion}

The polymer is considered as a differentiable space curve $\bm r(s, t)$ of length $L$, with contour coordinate $s$ ($-L/2 \leqslant s \leqslant L/2$) and time $t$. Activity is introduced in analogy to an active Ornstein-Uhlenbeck particle (AOUP)\cite{das:18.1} by assigning a propulsion velocity $\bm v (s,t)$  to every point $\bm r (s,t)$ (cf. Fig.~\ref{fig:sketch}) \cite{bech:16,eise:16,eise:17},  which changes in an independent manner. The equation of motion is then given by the Langevin equation \cite{harn:96,petr:06}
\begin{align}  \label{eq:langevin}
 & \frac{\partial \bm r(s,t)}{\partial t}  =  \   \bm v (s,t) + \int_{-L/2}^{L/2} \! ds' \,  \brh (\bm r(s),\bm r(s'))  \\ \nonumber & \times \left[ 2\nu k_BT \frac{\partial^2 \bm r (s',t )}{\partial s'^2}
   - \epsilon k_BT \frac{\partial^4 \bm r (s',t)}{\partial s'^4} + \bm \varGamma(s',t) \right]   \, ,
\end{align}
with boundary conditions for free ends as specified in Refs.~\onlinecite{harn:95,eise:16,eise:17}.
The tensor $\mathrm{\bf H}(\bm r(s),\bm r(s'))$ accounts for hydrodynamic interactions; it is defined as ${\bf
H}({\bm r}(s),{\bm r}(s')) = {\bm \Omega}({\bm r}(s)-{\bm r}(s')) +
{\bf I} \delta(s-s')/3 \pi \eta $, where the second term on the right hand side describes the local friction, and
\begin{align}
{\bm \Omega}(\Delta \bm r) = \frac{1}{8 \pi \eta | \Delta \bm r |} \left( {\bf I} + \frac{\Delta \bm r  \otimes \Delta \bm r }{|\Delta \bm r|^2} \right) \,
\end{align}
is the Oseen tensor \cite{doi:86,dhon:96}. The terms in Eq.~(\ref{eq:langevin}) with the second and forth derivative capture
chain flexibility, i.e., chain entropy, and bending forces, respectively. The Lagrangian multipliers $\nu (s)$ and $\nu_0=\nu(\pm L/2)$ account for the inextensibility of the polymer (we will denote $\nu$ as stretching coefficient in the following), and $\epsilon$ characterizes the bending stiffness \cite{wink:92,wink:03}. For a polymer in three dimensions, previous studies yield $\epsilon =
3/4p$ and $\nu_0=3/4$, where $p=1/2l_p$  and $l_p$ is the persistence
length \cite{wink:92,wink:03}. Adopting a mean-field approach, the stretching coefficient $\nu$ is independent of $s$ and is determined by the global constraint
\begin{align} \label{eq:constraint}
\int_{-L/2}^{L/2} \lla \left(\frac{\partial \bm r(s)}{\partial s}\right)^2 \rra d s = L .
\end{align}
The stochastic force ${\bm \varGamma} (s,t)$ is assumed to be
stationary, Markovian, and Gaussian. \cite{doi:86,risk:89}

Within the AOUP description of the analytical calculations, the active velocity $\bm v(s,t)$ is a non-Markovian but
Gaussian stochastic process with zero mean and the correlation function \cite{elge:15,wink:16,eise:16,sama:16}
\begin{align} \label{eq:corr_colored}
\lla \bm v(s,t) \cdot \bm v(s',t')\rra = v_0^2 l e^{-\gamma_R |t-t'|} \delta(s-s') \ .
\end{align}
Here, $v_0$ is the constant propulsion velocity and $\gamma_R$ characterizes the decay of the velocity correlation function. For a spherical colloid in solution, the relation $\gamma_R=2 D_R$ applies, where $D_R$ is the rotational diffusion coefficient. The correlation function (\ref{eq:corr_colored}) emerges due to a diffusive motion of either the Ornstein-Uhlenbeck process for the active velocity, or by the change of the propulsion direction (unit vector) of an ABP.\cite{eise:16,das:18.1}. Since only first and second moments of the active velocity are required for the current analytical studies, the results are independent of the underlying active velocity dynamics of an active site---either AOUP or ABP. Further details on the derivation of the equations of motion are presented in Ref.~\ocite{eise:16}, including a discussion of the factor $l$ in Eq.~(\ref{eq:corr_colored}).

Self-propelled systems are force and torque free \cite{elge:15,bech:16}. Hence, only conservative and random forces give rise to Stockeslet-type hydrodynamic interactions in Eq.~(\ref{eq:langevin}). However, we neglect force-dipole, source-dipole, and higher multipole flow field contributions, as they decay as ${\cal O}(r^{-2})$ with distance compared to a $1/r$ decay of the Stokeslet flow field \cite{lask:15}.

\subsection{Solution of the Equations of Motion}

\subsubsection{Hydrodynamic Tensor: Preaveraging Approximation}

The hydrodynamic tensor renders Eq.~(\ref{eq:langevin}) a nonlinear and nonlocal equation of motion. In order to obtain an (approximate) analytical solution, we apply the preaveraging approximation originally proposed by Zimm \cite{doi:86,zimm:56}, where the hydrodynamic tensor is replaced by its average over the stationary state distribution function, i.e., ${\bf H}(\bm r(s) - \bm r(s')) \rightarrow \lla{\bf H}(\bm r(s) - \bm r(s')) \rra = {\bf H}(s,s')$. Hence, Eq.~(\ref{eq:langevin}) turns into a linear equation---an Ornstein-Uhlenbeck process---with a Gaussian stationary-state distribution function for the distance $\Delta \bm r(s,s') = \bm r(s) - \bm r(s')$ of the form \cite{risk:89,das:18.1,harn:96,doi:86}
\begin{align} \label{eq:stat_state}
\Psi( \Delta \bm r ) = \left(\frac{3}{2\pi a^2(s,s') }\right)^{3/2}\exp \left(-\frac{3\Delta \bm r^2}{2 a^2(s,s')} \right) \, ,
\end{align}
with $a^2(s,s') = \lla (\bm r(s) - \bm r(s'))^2 \rra$. Note that $a(s,s')$ is not necessarily a function of $s-s'$ only. The explicit form of  $a(s,s)$ will be discussed  later. Preaveraging yields \cite{harn:96}
\begin{align} \label{eq:oseen_pre_avr}
{\bm \Omega}(s,s')
 = \frac{\Theta (|s-s'|-l}{3 \pi \eta} \sqrt{\frac{3}{2 \pi a^2(s,s')}} {\bf I} = \Omega (s,s') {\bf I} \, ,
\end{align}
and the hydrodynamic tensor becomes
\begin{align}
{\bf H}(s,s') =  \left[\frac{\delta(s-s')}{3\pi \eta} + \Omega(s,s') \right] {\bf I} = H(s,s') {\bf I} \, .
\end{align}
The Heaviside step function, $\Theta(x)$, in Eq.~(\ref{eq:oseen_pre_avr}) introduces a lower cut-off, which we choose as $l$.  \cite{harn:96} In a touching bead model of a polymer, $l$ is the bead diameter and, hence, the polymer thickness.

The preaveraging approximation has very successfully been applied to describe the dynamics of DNA \cite{petr:06} and semiflexible polymers. \cite{harn:96} Even quantitative agreement between analytical theory and simulations of the full hydrodynamic contribution of rather stiff polymers is achieved, \cite{hinc:09} as well as with measurements on DNA. \cite{petr:06,hinc:09.1} This demonstrates that preaveraging is also suitable for rather stretched polymers, it, however, fails for rodlike objects.\cite{wink:07.1}

\subsubsection{Eigenfunction Expansion}

The final linear equation is solved by the eigenfunction expansion
\begin{align} \label{eq:eigen_expand}
\bm r(s,t) = \sum_{n=0}^{\infty} \bm \chi_n (t) \vphi_n(s)
\end{align}
in terms of the eigenfunctions $\vphi_n$ of the equation
\begin{align}
\epsilon k_BT \frac{d^4}{ds^4} \vphi_n(s) - 2\nu k_BT \frac{d^2}{ds^2} \vphi_n(s) = \xi_n \vphi_n(s) \, ,
\end{align}
with the eigenvalues ($n \in \mathbb{N}_{0}$)
\begin{align} \label{eq:eigenvalue}
\xi_n = k_BT \left(\epsilon \zeta_n^4+ 2\nu \zeta_n^2 \right) .
\end{align}
The mode numbers $\zeta_n$ follow from the boundary conditions. The respective eigenfunctions and eigenvalues are explicitly presented in Refs.~\ocite{harn:95,eise:16,eise:17}.
Specifically, in the limit of a flexible polymer, $pL \gg 1$, the eigenfunctions are
\begin{align}
\varphi_0 & = \sqrt{\frac{1}{L}} \, ,\\
\varphi_n(s) & = \sqrt{\frac{2}{L}} \sin \left( \frac{n\pi s}{L} \right)\, , \, \forall n \ \text{odd}\\
\varphi_n(s) & = \sqrt{\frac{2}{L}} \cos \left( \frac{n\pi s}{L} \right)\, , \, \forall n \  \text{even} \, ,
\end{align}
with the wave numbers $\zeta_n = n \pi/L$ and the eigenvalues $\xi_n = 2 \nu k_BT \pi^2 n^2/L^2$.

The equation of motion \eqref{eq:langevin}  yields the  Langevin equation for the mode amplitudes, $\bm \chi_n(t)$,
\begin{align} \label{eq:chi_eq}
\frac{d  {\bm \chi}_m (t)}{d t } = \sum_{n=0}^{\infty}  H_{mn}\left[\bm \varGamma_n (t) - \xi_n  {\bm \chi}_n (t) \right] + \bm v_m(t) .
\end{align}
The mode representation of the hydrodynamic tensor is $H_{nm} = (\delta_{nm} + 3 \pi \eta \Omega_{nm} )/3\pi \eta$, with the preaveraged Oseen tensor $\Omega_{nm}$.\cite{harn:96}
The second moments of the stochastic-force amplitudes $\bm \varGamma_n$ are given by
\begin{align} \label{eq:corr_stoch_mode}
\langle  \varGamma_{n \alpha}(t) \, \varGamma_{m \beta}(t')  \rangle & = \, 2 k_B T \delta_{\alpha \beta } \delta(t-t') \, H_{nm}^{-1} \, ,
\end{align}
with $\alpha,\beta \in \left \{ x,y,z \right \}$.  The mode representation of the correlation function (\ref{eq:corr_colored}) of the active velocity is \cite{eise:16}
\begin{align} \label{eq:mode_corr_v}
\langle \bm v_n(t) \cdot \bm v_m (t') \rangle = \, v_0^2 l e^{-\gamma_R |t-t'|} \delta_{nm} \, .
\end{align}

In Eq.~(\ref{eq:chi_eq}), all modes are coupled in general and the set of equations can only be solved numerically. To arrive at an analytical solution, we neglect the off-diagonal terms of the hydrodynamic mode tensor $H_{nm}$, which leads to the decoupled equations  \cite{doi:86,harn:96,petr:06}
\begin{equation} \label{eq:chi_eqMotHydUncoup}
\frac{d \bm \chi_n (t)}{d t } = - \frac{1}{\tilde \tau_n} \bm \chi_n + H_{nn} \bm \varGamma_n (t)  + \bm v_n(t) \, ,
\end{equation}
with the relaxation times
\begin{align} \label{eq:relax_time}
\tilde{\tau}_n = \frac{1}{H_{nn} \xi_n} = \frac{\tau_n}{1 + 3 \pi \eta \Omega_{nn}} ,
\end{align}
and $\tau_n=3 \pi \eta/\xi_n$,  the relaxation times in absence of hydrodynamic interactions; for flexible polymers $\tau_n= 3 \eta L^2/(2 \nu k_BT \pi n^2)$. \cite{harn:95,eise:17}

The stationary-state solution of Eq.~(\ref{eq:chi_eq}) is
\begin{align} \label{eq:chi_n}
 \bm \chi_n (t) & = \int_{-\infty }^{t} \! dt' \, e^{-(t-t') / \tilde{\tau}_n }  \left[\bm v_n(t') +  H_{nn} \bm \varGamma_n (t') \right]  \,
 \end{align}
for $n>0$, and for  $n=0$ the solution is
\begin{align} \label{eq:chi_0}
 \bm \chi_0 (t) & = \bm \chi_0(0) + \int_{0 }^{t} \! dt' \, \left[\bm v_{(0)}(t')+  H_{00} \bm \varGamma_0 (t')  \right] \, .
 \end{align}

 \subsubsection{Averages and Correlation Functions}

With the correlation functions of the stochastic forces (\ref{eq:corr_stoch_mode}) and velocities (\ref{eq:mode_corr_v}), the correlation functions of the mode amplitudes become
($t \geq t'$)
\begin{align} \label{eq:corr_mod_n}
 \lla  \bm \chi_n(t)  \cdot   \bm \chi_m (t') \rra =  & \  \delta_{nm} \left( \frac{ k_BT \tau_n}{\pi \eta } e^{-|t-t'|/\tilde{\tau}_n } \right. \\ \nonumber & \hspace*{-2cm}
 \left. + \frac{v_0^2 l {\tilde{\tau}}_n ^2}{1-(\gamma_R \tilde{\tau}_n)^2} \left[e^{-\gamma_R |t-t'|} - \gamma_R \tilde{\tau}_n e^{-|t-t'|/ \tilde{\tau}_n} \right] \right) \, , \\ \label{eq:coor_mod_0}
 \lla  \bm \chi_0(t) \cdot  \bm \chi_0(t')\rra  =  & \,  \lla \bm \chi_0^2 (0) \rra  + 6 k_B T \, H_{00} \, t' \\ \nonumber  & \hspace*{-2cm}
 + \frac{v_0^2 l}{\gamma_R^2}\left[ 2 \gamma_R t' -1 -e^{\gamma_R (t'-t)} + e^{-\gamma_R t} + e^{-\gamma_R t'} \right] \, .
\end{align}

The eigenfunction expansion (\ref{eq:eigen_expand}) and the  correlation functions (\ref{eq:corr_mod_n}) permit us to calculate the mean square distance $a^2(s,s')$. Explicitly, we find
\begin{align} \label{eq:msqd}
a^2(s,s') = \sum_{n=1}^{\infty} \lla \bm \chi_n^2\rra (\vphi_n(s) - \vphi_n(s'))^2 ,
\end{align}
with the stationary-state average
\begin{align} \label{eq:chi_st_st}
\lla \bm \chi_n^2 \rra = \frac{k_BT \tau_n}{\pi \eta} + \frac{v_0^2l \tilde \tau_n^2}{1+\gamma_R \tilde \tau_n} .
\end{align}

As for passive polymers, the relaxation behavior  \eqref{eq:corr_mod_n} is determined by hydrodynamics. Remarkably, however, the mode-amplitude correlation functions \eqref{eq:chi_st_st} depend on the hydrodynamic interactions via the relaxation times \eqref{eq:relax_time}. Thus, HI changes, additionally to the dynamics, also the stationary-state conformational properties of an active polymer.

\subsubsection{Hydrodynamic Tensor: Mode Representation}

In order to determine the relaxation times $\tilde \tau_n$, the double integral
\begin{align} \label{eq:oseen_mode_dist}
 \Omega_{nn}  = \sqrt{\frac{1}{6 \pi^3 \eta^2}}  \int_{-L/2 }^{L/2} \int_{-L/2 }^{L/2}   \Theta (|s-s'|-l_{c})  \frac{\varphi_n(s) \varphi_n(s')}{\sqrt{a^2(s,s') }} ds' ds
\end{align}
needs to be evaluated,
which itself depends via $a^2(s,s')$ on the Oseen tensor $\Omega_{nn}$. Hence,  the equation has to be solved in an iterative and self-consistent manner, where the double integration, combined with the summation of Eq.~(\ref{eq:msqd}),  constitutes a major computational challenge. To arrive at a more easily tractable expression with a single integral, we apply standard approximations for the integrals over the functions $\vphi_n$ in Eq.~(\ref{eq:oseen_mode_dist}) as, e.g.,  described in Ref.~\ocite{doi:86} for  a flexible polymer. For a passive semiflexible polymer,  $a^2(s,s')$ is only a function of the difference $|s-s'|$ \cite{wink:94,harn:96}. This is no longer the case  in the presence of activity, where  $a^2(s,s')$ depends on $s-s'$ and $s+s'$ in general. In fact, an analytical expression of $a^2$ for a flexible ABPO-HI can be calculated by performing the sum in Eq.~(\ref{eq:msqd}).  To obtain an approximate expression, which depends on the difference $|s-s'|$ only, we replace the difference of the eigenfunctions in Eq.~(\ref{eq:msqd}) by the expression approximately valid for a passive polymer, namely $\vphi_n(s)- \vphi_n(s')=2 \sin(n\pi(s-s')/2L)$ for $n$ odd, and $\vphi_n(s)- \vphi_n(s')=0$ for $n$ even.
As a result, we obtain
\begin{align} \label{eq:msqd_approx}
a^2(s)= \frac{8}{L} \sum_{n, \text{odd} } \left( \frac{ k_BT \tau_n}{\pi \eta }    + \frac{v_0^2 l \tilde{\tau}_n ^2}{1+\gamma_R \tilde{\tau}_n} \right) \sin^2\left(\frac{n\pi }{2L} s \right)  \, ,
\end{align}
and, hence, $\Omega_{nn}$ is given by
\begin{align} \label{eq:omega_single}
\Omega_{nn} =
\sqrt{\frac{2}{3\pi^3}} \frac{1}{\eta L}  \int_{l_{c} }^{L} \! \frac{L-s}{\sqrt{a^2(s)} } \cos \left( \frac{n\pi }{L} s \right) \, ds .
\end{align}
Aside from the distance $a^2(s-s')$, which depends on activity via the relaxation times,  this expression is identical with that of a passive polymer. \cite{harn:96}  As shown in Sec.~S-I of the ESI, the approximations of Eqs.~\eqref{eq:msqd_approx} capture the dependence of $a^2$ on the contour coordinate well, the better the larger the P\'eclet number.

Assuming a linear dependence of $a^2(s-s')$ on $|s-s'|$, i.e., $a^2(s) = a_0^2|s| L$, as for a passive flexible polymer \cite{harn:96}, we obtain the analytical solution of Eq.~\eqref{eq:omega_single},
\begin{align} \label{eq:omega_app}
\Omega_{nn} = \frac{1}{\sqrt{3 \pi^3} \eta a_0} \frac{1}{\sqrt{n}} ,
\end{align}
in analogy to the Zimm approach \cite{zimm:56,doi:86,harn:96}. Choosing for $a_0^2$ the result of a flexible ABPO-HI, namely $a_0^2 = 1 /\mu pL +Pe^2 /6 \mu p L \Delta$, where the P\'eclet number $Pe$ and $\Delta$ are defined in Eq.~\ref{eq:peclet}, and $\mu$ is given by $\mu = 2 \nu/(3 p)$, \cite{eise:16}
we obtain
\begin{align} \label{eq:omega_limit}
\Omega_{nn} \sim \frac{\sqrt{pL \mu}}{Pe \sqrt{n}}
\end{align}
for $Pe \gg1$. \\

In the following, when not indicated otherwise, the approximate expressions \eqref{eq:msqd_approx} and \eqref{eq:omega_single} are used for the calculation of the Oseen tensor. Moreover, we use $\Delta = 1/3$, the value of a spherical colloid of diameter $l$ in solution.

\subsection{Stretching Coefficient and Relaxation Times} \label{sec:relax_time}

The stretching coefficient and relaxation times are interdependent and need to be determined simultaneously. Due to  nonlinearities, specifically in $\Omega_{nn}$, the respective quantities can only be determined numerically.

We  focus here on flexible polymers, where $pL \gg1$, and we set $L/l=pL$.  Then, in terms of the eigenfunction expansion \eqref{eq:eigen_expand}, Eq.~\eqref{eq:constraint} for the stretching coefficient $\nu$, respectively $\mu = 2 \nu/3p$, becomes
\begin{align} \label{eq:constraint_mode}
\sum_{n=1}^{\infty} \left[ \frac{k_BT \tau_n}{\pi \eta} + \frac{v_0^2l \tilde \tau_n^2}{1+\gamma_R \tilde \tau_n} \right] \zeta_n^2 = L ,
\end{align}
with the relaxation times (Eq.~\eqref{eq:relax_time})
\begin{equation} \label{eq:tau_asym_flex}
\tilde{\tau}_n =  \frac{\tau_R}{\mu n^2 (1+3\pi \eta \Omega_{nn})} \ ,
\end{equation}
where $\tau_R \equiv \eta L^2/(\pi k_B T p)$  is the  Rouse relaxation time \cite{harn:95,doi:86}.
In the non-hydrodynamic case, i.e., $\Omega_{nn}=0$, we find the asymptotic solution  $\mu = Pe^{4/3}/6\Delta$ of Eq.~\eqref{eq:constraint_mode}, independent of $pL$.\cite{eise:16} For the relaxation times, we recover the Zimm behavior $\tilde{\tau}_n = \tau_Z / n^{3/2}$ at $Pe=0$, with the longest relaxation time $\tau_Z = \eta (L/p)^{3/2} /(\sqrt{3\pi} k_B T)$ (Zimm relaxation time).

\begin{figure}[t!]
\begin{center}
\includegraphics[width=\columnwidth]{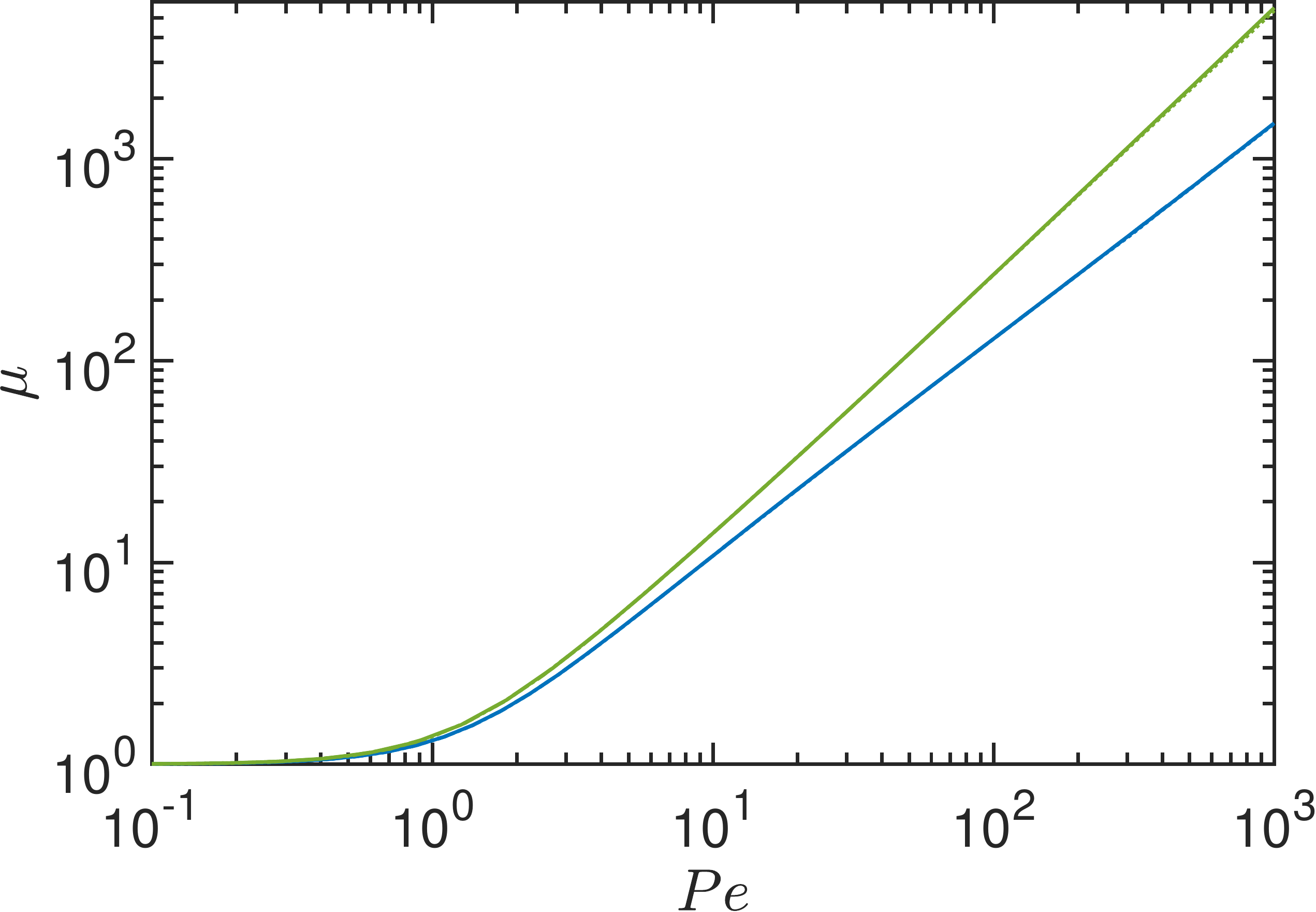}
\end{center}
\caption{Normalized stretching coefficient $\mu=2\nu /(3p)$, solution of Eq.~\eqref{eq:constraint_mode}, as function of the P\'eclet number $Pe$ for flexible polymers with $pL=10^3$, $10^4$, and $10^5$ without HI (green) and with HI (blue). The results are independent of polymer length.
 } \label{fig:lag}
\end{figure}

The scaled stretching coefficient, $\mu$, is presented in Fig.~\ref{fig:lag} as a function of the  P\'eclet number. For the considered polymer lengths and stiffness, $\mu$ is independent of $pL$. Moreover, it increases approximately linearly with P\'eclet number for $Pe \gtrsim 5$, somewhat weaker than $\mu$ of comparable passive polymers. Hence, hydrodynamics modifies the stretching coefficient.

\begin{figure}[t]
\begin{center}
\includegraphics[width=\columnwidth]{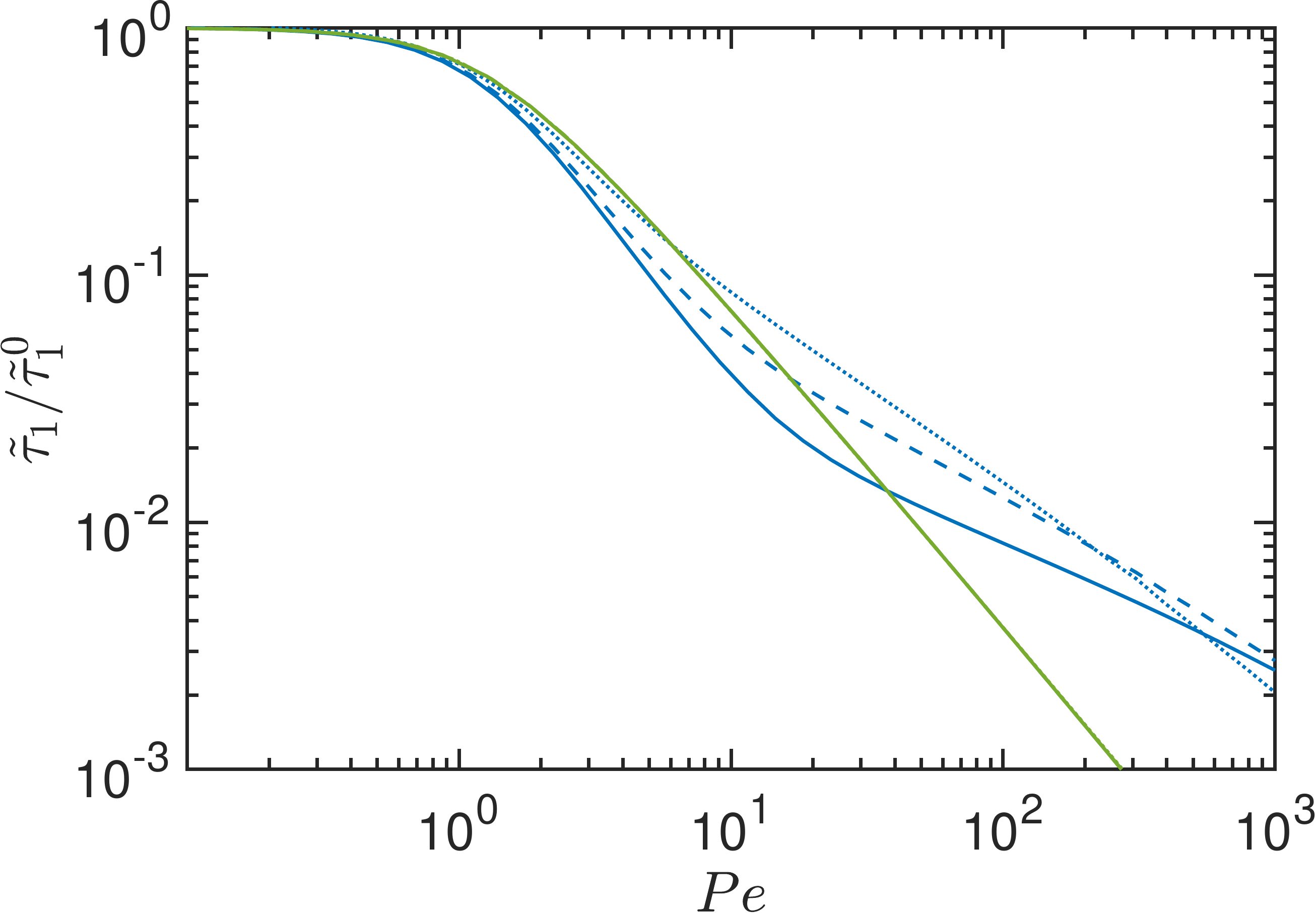}
\end{center}
\caption{The longest polymer relaxation time $\tilde{\tau}_1$, Eq.~\eqref{eq:tau_asym_flex}, normalized by the corresponding passive value  $\tilde{\tau}_1^0$ as function of the P\'eclet number $Pe$ for flexible polymers with $pL=10^5$ (solid), $10^4$ (dashed), and $10^3$ (dotted). The green line shows the result of an active polymer in absence of HI, where $\tau_1 \sim Pe^{-4/3}$.} \label{fig:tau1}
\end{figure}

Figure~\ref{fig:tau1} displays  longest relaxations times, $\tilde \tau_1$, as function of $Pe$. For $1 < Pe \lesssim 10$, hydrodynamic interactions enhance the decay of the relaxation time with increasing $Pe$ compared to the non-hydrodynamic case, specifically for  $pL > 10^3$. Note that $\mu \sim Pe$, independent of the polymer length in the considered length regime. This is a consequence of an increase of $\Omega_{nn}$ with increasing $Pe$ (cf. Sec. S-I. of ESI). In contrast, a slower decay of $\tilde \tau_1$ is obtained for $Pe >10$. Here, we find a strong polymer-length dependence, which is related to particular values of $\Omega_{nn}$ (cf. Fig.~S.2). The approximation \eqref{eq:omega_limit} yields the relation $\tilde \tau_1 \sim 1/\sqrt{Pe}$ for $3 \pi \eta \Omega_{nn} \gg 1$, which describes the P\'eclet number dependence well in the interval $10< Pe <10^3$ for $pL=10^5$. As shown in Fig.~S.2 for shorter polymers,  $\Omega_{nn}$ varies more  slowly with $Pe$ and, hence, $\tilde \tau_1$ decays faster with increasing $Pe$. In the limit $Pe \to \infty $, $\Omega_{nn}$ becomes very small and the contribution to the relaxation times vanishes gradually. Hence, $\tilde \tau_1$ approaches the asymptotic dependence $\tilde \tau_1 \sim 1/Pe$, determined by $\mu$.

Results on the mode-number dependence of the relaxations times for various $Pe$ are presented in Sec.~S-III of the ESI.
The intricate dependence of $\Omega_{nn}$ on the relaxation times poses a  major challenge for an (approximate) analytical solution, a problem we were not able to overcome so far.

\subsection{Conformational Properties} \label{sec:conf}

\begin{figure}[t]
\begin{center}
\includegraphics[width=\columnwidth]{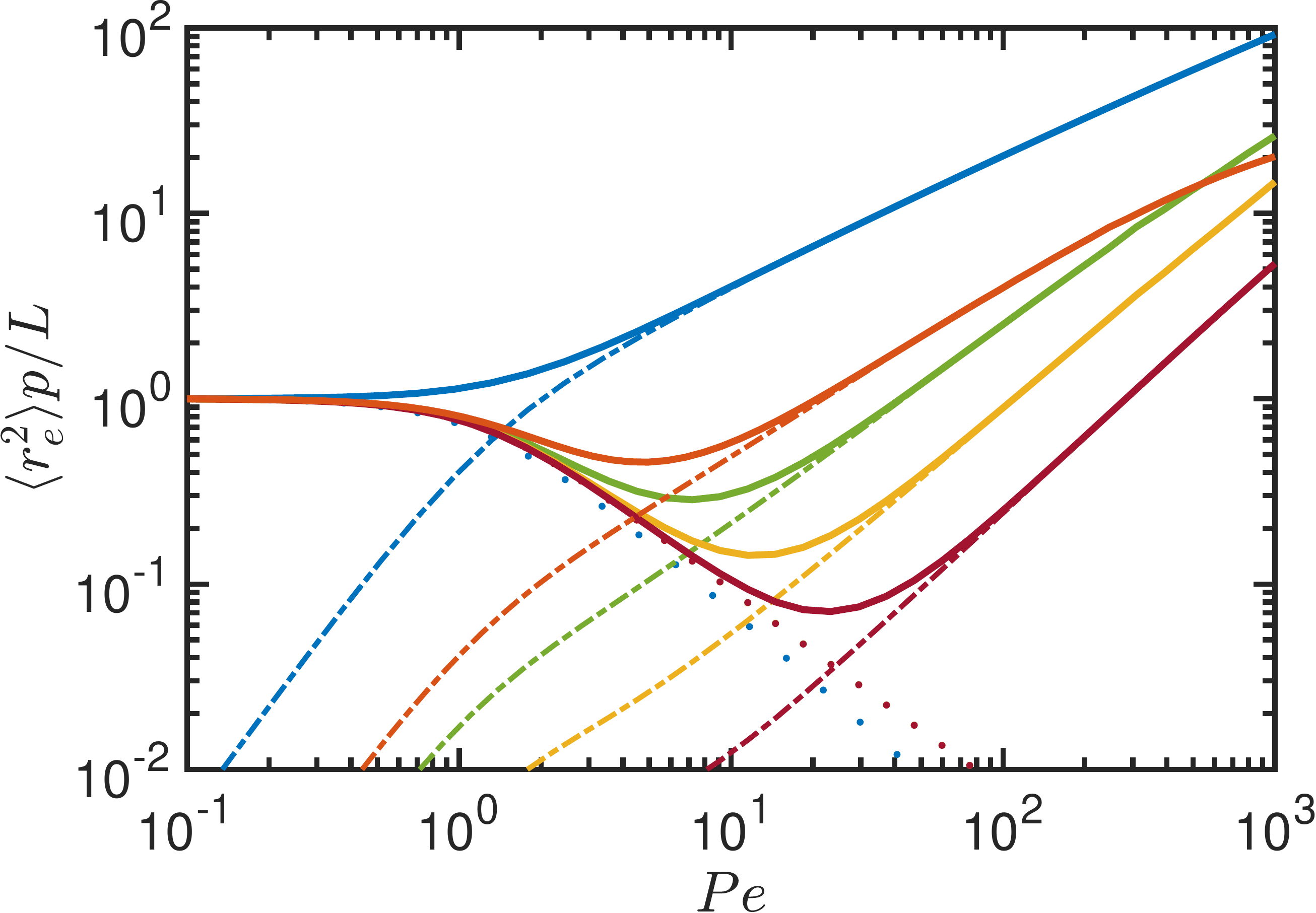}
\end{center}
\caption{Polymer mean square end-to-end distance $\langle \bm r_e^2 \rangle$, Eq. ~\eqref{eq:end-to-end}, as a function of the P\'eclet number $Pe$ for flexible ABPOs+HI of length $pL = 2\times 10^2$ (orange), $10^3$ (green), $10^4$ (yellow), and $10^5$ (magenta). The blue lines correspond to a free-draining flexible polymer with $pL=50$.  The dotted curves represent the contribution with the relaxation times $\tau_n$ and the dashed-dotted curves that with  $v_0^2$ of Eq.~\eqref{eq:end-to-end}, respectively.
 } \label{fig:end}
\end{figure}

The conformational properties of a polymer are characterized by the  mean square end-to-end distance $\lla \bm r_e ^2\rra = \lla (\bm r(L/2) - \bm r(-L/2))^2 \rra$, which is given by
\begin{align} \label{eq:end-to-end}
\lla \bm r_e^2\rra = \frac{8}{L} \sum_{n, \, \text{odd}} \left( \frac{k_BT \tau_n}{\pi \eta} + \frac{v_0^2l \tilde \tau_n^2}{1+\gamma_R \tilde \tau_n} \right)
\end{align}
in terms of the mode amplitudes of Eq.~\eqref{eq:chi_st_st}. Numerical results for $\lla \bm r_e^2\rra$ of flexible polymers  ($pL \gg 1$) are shown in Fig.~\ref{fig:end} for various polymer lengths. Starting from the equilibrium value $\lla \bm r_e^2\rra = L/p$ at $Pe=0$, ABPOs+HI  first shrink with increasing activity and then swell for higher $Pe$ (solid lines), in qualitative agreement with the simulation results of Sec.~\ref{sec:sim_conformations}. In the asymptotic limit $Pe \to \infty$, a limiting value $ \lla \bm r_e^2\rra < L^2$ is assumed.  Thereby, the shrinkage strongly depends on the polymer length and is more pronounced for longer polymers. As shown in Fig.~\ref{fig:end}, flexible ABPO-HI exhibit a drastically different behavior and swell monotonically with increasing activity. The reason for the qualitatively different conformational properties rests on the different polymer-length dependence of the Rouse and Zimm relaxation times, where $\tau_R/\tau_Z \approx \sqrt{pL}$. Hence,  in the presence of hydrodynamic interactions,  relaxation times are shorter by the factor $1/\sqrt{pL}$, which can be orders of magnitude for long flexible polymers.

Hydrodynamic interactions lead to a polymer-length dependence of the swelling with increasing P\'eclet number ($Pe \ll \infty$), as shown in Fig.~\ref{fig:end}. For polymer lengths in the range  $pL \approx 10^2 - 10^3$, $\Omega_{nn}$ depends only weakly on the mode number (cf. Fig.~S.2), hence, replacement of the relaxation times $\tilde \tau_n$ by the relaxation times $\tau_Z/n^{3/2}$ yields
\begin{align} \label{eq:end-to-end_zimm}
\lla \bm r_e^2\rra = \frac{L}{p\mu} \left[1 + \frac{4 Pe^2}{(3 \pi)^{3/2} \Delta \sqrt{pL}} \left(1-\frac{1}{2\sqrt{2}} \zeta(3/2) \right) \right] ,
\end{align}
in the limit $\gamma_R \ttau_n \gg 1$, where $\zeta(x)$ is Riemann's zeta function ($\zeta(3/2) \approx 2.61$).
In the limit of very large $pL$ and $Pe \gg 1$, at least in the vicinity of $pL =10^5$, $\ttau_n$ can be approximated by $\Omega_{nn}$ of Eq.~\eqref{eq:omega_limit}, which yields
\begin{align} \label{eq:end-to-end_limit}
\lla \bm r_e^2 \rra \sim \frac{L}{p \mu} \frac{Pe^3}{\sqrt{\mu pL}} .
\end{align}
Thus, we find the same dependence on $pL$ for both, small and large $pL$, and $\langle \bm r_e^2 \rangle/L^2$ decreases as $1/(pL)^{3/2}$. The dependence on $Pe$ changes from $\lla \bm r_e^2 \rra \sim Pe$ for $pL\approx 10^2 - 10^3$ to  $\langle \bm r_e^2 \rangle \sim Pe^{3/2}$ for $pL \approx 10^5$, because $\mu \approx Pe$. For an ABPO-HI, we found instead $\langle \bm r_e^2 \rangle \sim L Pe^{2/3}/p$, \cite{eise:16}
since for such a polymer $\mu \sim Pe^{4/3}$. Hence, hydrodynamic interactions lead to a qualitative different $Pe$ dependence.

Figure  \ref{fig:end} shows the individual contributions to $\lla \bm r_e^2 \rra$---the term with the relaxation times $\tau_n$ (dotted lines) and that with $v_0^2$ (dashed-dotted lines) in Eq.~\eqref{eq:end-to-end}, respectively. The initial shrinkage of $\langle \bm r_e^2 \rangle$ with increasing $Pe$ is caused by the decreasing relaxation times $\tau_n \sim 1/\mu$ with increasing activity.  In the thermal contribution of Eq.~\eqref{eq:end-to-end_zimm},  $\langle \bm r_e^2 \rangle \sim L/(p\mu)$,  the stretching coefficient $\mu$ increases with increasing activity, which formally implies a decreasing persistence length below the value of a passive polymer, corresponding to more compact conformations than of the passive case. The $v_0^2$-dependent term causes a swelling of the polymer. For an ABPO-HI, the competing effects lead to an overall swelling, since swelling exceeds shrinkage. In case of an ABPO+HI, swelling is weaker due to fluid-induced collective motion (cf. Fig.~\ref{fig:sketch}) compared to the random motion of an ABPO-HI, and  $\langle \bm r_e^2 \rangle $ assumes a minimum.   Mathematically, this is reflected by the shorter relaxation times $\tilde \tau_n$ compared to $\tau_n$.  Hydrodynamic interactions accelerate the polymer dynamics and higher $Pe$ are required to achieve a significant swelling of an ABPO+HI.

The exponents of the (approximate) power-law regimes for the various $pL$ values approximately exhibit the above predicted scaling relations with respect to $Pe$ ($Pe>10$). The shift of the dashed-dotted curves in Fig.~\ref{fig:end} to smaller $\lla \bm r_e^2\rra$  with increasing polymer length, $pL$, reflects the discussed decrease in relaxation times by hydrodynamics.

\begin{figure}[t!]
\begin{center}
\includegraphics[width=\columnwidth]{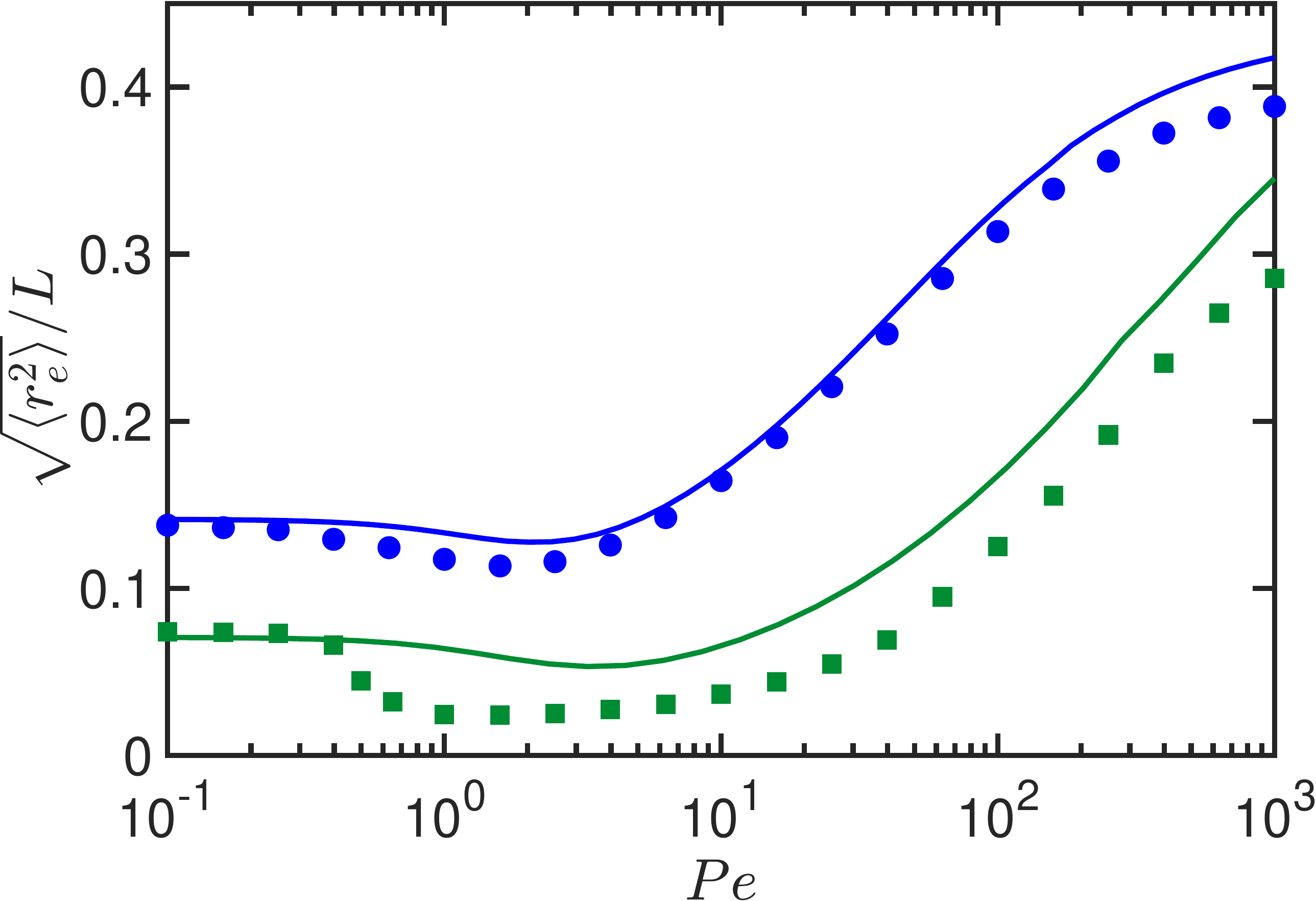}
\end{center}
\caption{Mean square end-to-end distance as a function of the P\'eclet number of flexible ABPOs+HI  for the monomer number   $N_m=50$ ($pL=50$) (blue) and $N_m=200$ ($pL=200$) (green). Solid lines correspond to analytical and symbols to simulation results.} \label{fig:comp_sim_theory}
\end{figure}

Figure~\ref{fig:comp_sim_theory} shows a comparison of analytical and simulation results. We find good agreement for short polymers ($N_m=50$), but theory yields a  less pronounced shrinkage for the longer polymers. 
We like to emphasize that for an ABPO-HI the theoretical approach reproduces the simulation data very well. \cite{eise:17.1} The reason of the discrepancy is not evident, but is related to the applied approximations, which seem to underestimate hydrodynamic effects. We speculate that the preaveraging approximation may fail, because active fluctuations could be large and the replacement of $\mathrm{\bf H}(\bm r - \bm r')$ by $\langle \mathrm{\bf H (\bm r - \bm r')} \rangle$ no longer be justified.
Yet, the analytical expression captures the qualitative behavior, and even more quantitatively the swelling behavior at large $Pe$ is reasonably well reproduced, although the asymptotic value for $Pe\to \infty$ is somewhat overestimated due to the applied mean-field approximation of the bond-length constraint.

\subsection{Dynamical Properties} \label{sec:dyn}


The dynamics of the polymers is characterized by the site mean square displacement (MSD) averaged over the polymer contour, $\langle \lbar{\Delta \bm r^2}(t) \rangle = \int \langle (\bm r(s,t) - \bm r(s,0))^2 \rangle ds/L$, which yields
\begin{align}   \label{eq:meansquaredisp_SP}
& \lla  \lbar{\Delta  \bm r^2}(t) \rra  = \lla \Delta  \bm r_{cm}^2(t) \rra +\lla \lbar{\Delta  \bm r_{0}^2}(t) \rra + \lla \lbar{\Delta  \bm r_{a}^2}(t) \rra    ,
\end{align}
with the center-of-mass mean square displacement
\begin{equation} \label{eq:amp_CMmsd_SP}
\lla \Delta  \bm r_{cm}^2(t) \rra  = \frac{6k_BT }{L} H_{00} t + \frac{2 v_0^2 l }{\gamma_R^2 L}\left( \gamma_R t -1 + e^{-\gamma_R t} \right) ,
\end{equation}
$H_{00}=(1+3\pi \eta \Omega_{00})/(3\pi \eta)$, the activity-modified equilibrium internal-dynamics contribution
\begin{align}  \label{eq:meansquaredisp_eq}
\lla  \lbar{\Delta  \bm r^2_0}(t) \rra  =  \frac{1}{L}\sum_{n=1}^{\infty}   \frac{6 k_BT \tau_n}{\gamma} \left(1 - e^{-t/\ttau_n}\right)  \ ,
\end{align}
and the active contribution
\begin{align} \label{eq:meansquaredisp_ac}
\lla  \lbar{\Delta  \bm r^2_a}(t) \rra  =  \frac{1}{L}\sum_{n=1}^{\infty}  \frac{2 v_0^2 l \ttau_n ^2}{1+ \gamma_R \ttau_n } \left( 1 - \frac{e^{-\gamma_R t} - \gamma_R \ttau_n e^{-t/\ttau_n}  }{1 - \gamma_R \ttau_n}  \right) .   \end{align}

Remarkably, in the center-of-mass MSD only the thermal contribution includes hydrodynamics, via $H_{00}$ \cite{petr:06}, which depends on activity through $\mu$, whereas the active term is identical with that of an ABPO-HI \cite{wink:16,eise:16,eise:17}. The reason is that swimming is force free and no Stokeslet is present.
Within the approximation $a^2(s,s') \approx a^2(s-s')$,  Eq.~\eqref{eq:omega_single} yields
\begin{align}
\Omega_{00} = \frac{8}{3 \sqrt{6 \pi^3} \eta a_0} .
\end{align}
Hence, for $3 \pi \eta \Omega_{00} \gg 1$, the thermal center-of-mass diffusion coefficient $D_0 =  k_BT \Omega_{00}/L$ increases somewhat due to activity in the range $1 \lesssim Pe \lesssim 50$, and decreases for higher $Pe$ (cf. Fig.~S.2).

In the asymptotic limit $Pe \to \infty$, the active polymer is  stretched, and the hydrodynamic contribution to thermal diffusion decreases (cf. Fig.~S.2 of ESI). As for a passive, rodlike polymer \cite{wink:07.1,dhon:96,doi:86}, asymptotically hydrodynamic interactions yield only small corrections with respect to the polymer-length dependence of a non-hydrodynamic (free-draining) polymer.

\begin{figure}[t!]
\begin{center}
\includegraphics[width=\columnwidth]{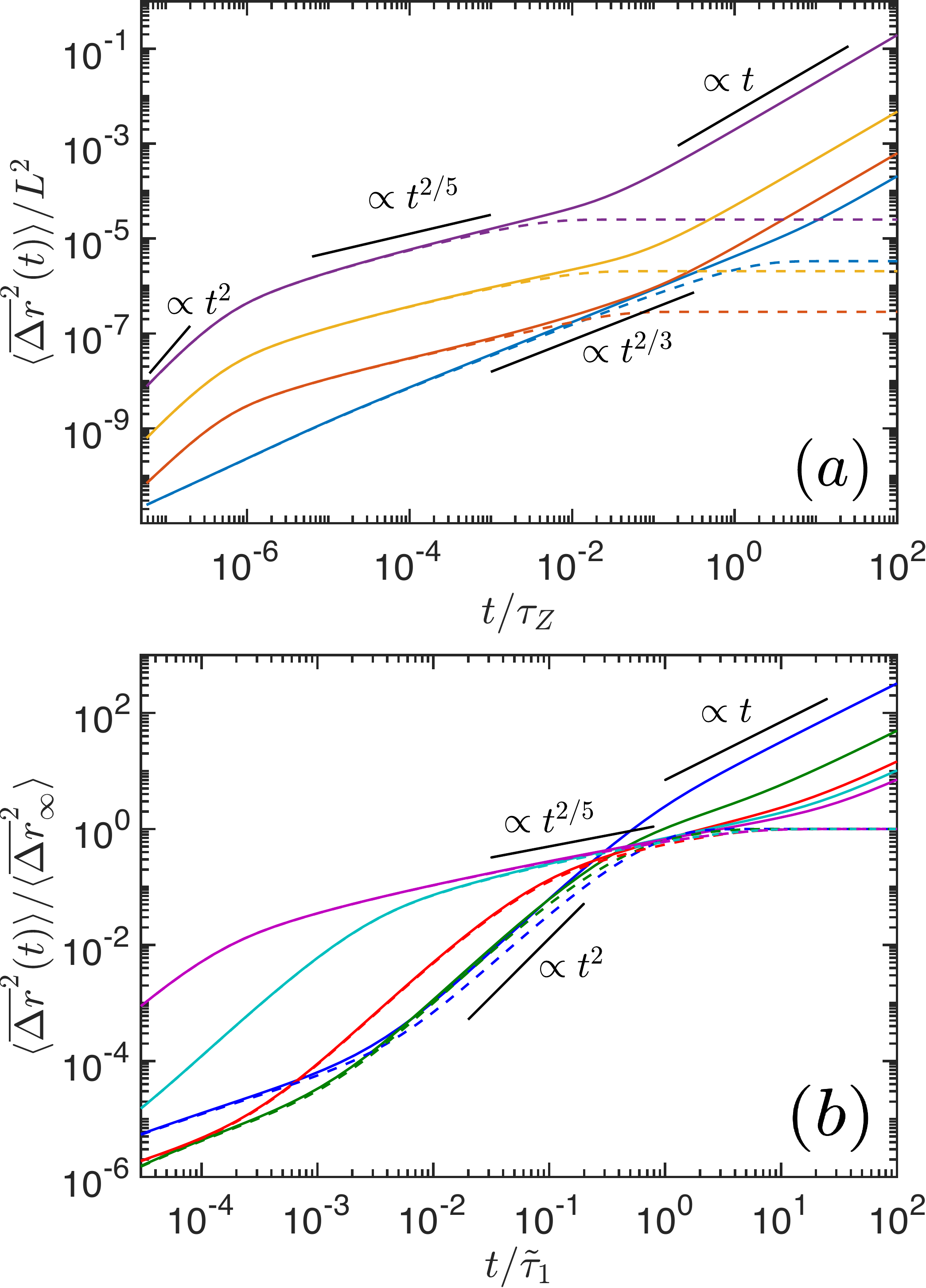}
\end{center}
\caption{Mean square displacement of flexible ABPO+HI, Eq.~\eqref{eq:meansquaredisp_SP}. (a)  MSDs for the P\'eclet numbers  $Pe = 10^{-2}$ (blue), $Pe = 2 \times 10^1$ (orange), $Pe = 1.5\times 10^{2}$ (yellow), and $Pe = 10^{3}$ (purple); the polymer  length is $pL = 10^5$. The time is scaled by the Zimm time $\tau_Z$ of a passive polymer. (b) MSDs for the polymer lengths $pL = 5 \times 10^1$ (blue), $2\times 10^2$ (green), $10^3$ (red), $10^4$ (cyan), and $10^5$ (purple) and $Pe = 10^2$. $\langle  \lbar{\Delta  \bm r^2}_{\infty} \rangle = 2 \langle \bm r_g^2 \rangle$ denotes the asymptotic value of the MSD in the center-of-mass reference.
The dashed lines correspond to the MSD in the polymer center-of-mass reference frame, Eqs.~\eqref{eq:meansquaredisp_eq} + \eqref{eq:meansquaredisp_ac}, and the solid lines to the overall MSD, Eq.~\eqref{eq:meansquaredisp_SP}. The black lines indicate power laws in the respective regimes.
} \label{fig:seg_msd}
\end{figure}

Figure~\ref{fig:seg_msd} displays the average site mean square displacement for various P\'eclet numbers.  For a passive flexible polymer, we recover the well-known Zimm behavior, with  $\langle \overline{\Delta  \bm r^2_0}(t) \rangle \sim t^{2/3}$ for $t/ \tau_Z \ll 1$, and a crossover to free diffusion for $t/\tau_Z \gg1$ \cite{doi:86,zimm:56,petr:06}. In the presence of activity, four time regimes can be identified
\begin{itemize}[leftmargin=*]
\item $t \to \infty$ --- The MSD is dominated by the linear time dependence of the center-of-mass dynamics, with the diffusion coefficient
\begin{align}
D = \frac{k_BT H_{00}}{L} +  \frac{v_0^2l}{3 \gamma_R L} .
\end{align}
The other terms approach a constant value equal to $2 \langle \bm r^2_g \rangle$, where $\langle \bm r^2_g \rangle$ is the active-polymer radius of gyration. The simulations results are in agreement with the active contribution to $D$.
\item $t/\tilde \tau_1 \ll 1 \ll \gamma_R t$ --- The active contribution to the MSD is dominated by ($\gamma_R \tilde \tau_1 \gg1 $)
\begin{align}
\lla  \lbar{\Delta  \bm r^2_a}(t) \rra  =   \frac{2 v_0^2 l}{\gamma_R L}\sum_{n=1}^{\infty}   \ttau_n  \left( 1 - e^{-t/\ttau_n}  \right)  .
\end{align}
With the power-law dependence $\tilde \tau_n = \ttone/n^{\tilde \alpha}$ of the relaxation time and by replacing the sum by an integral, we find
\begin{align}  \label{eq:msd_active}
\lla  \lbar{\Delta  \bm r^2_a}(t) \rra  =
 \frac{2 v_0^2 l \ttone}{\gamma_R L}  \left(\frac{t}{\ttone} \right)^{1-1/\tilde \alpha}  \int_0^{\infty} dx \ \frac{1-e^{ -x^{\tilde \alpha}}}{x^{\tilde \alpha}}  .
\end{align}
With the assumption $3 \pi \eta \Omega_{11} \gg 1$ and Eq.~\eqref{eq:omega_app}, which corresponds to the exponent $\tilde \alpha = 3/2$, Eq.~\eqref{eq:msd_active} yields
\begin{align} \label{eq:power_law}
\lla  \lbar{\Delta  \bm r^2_a}(t) \rra \sim \frac{L^2 Pe^{5/3}}{(pL)^2} t^{1/3} ,
\end{align}
i.e.,  a site sub-diffusive MSD dominated by the internal polymer dynamics. The exponent $\alpha'= 1-1/\tilde \alpha = 1/3$ of Eq.~\eqref{eq:power_law} approximately agrees with the full numerical result $2/5$ (cf. Fig.~\ref{fig:seg_msd}). Note that the exponent $\tilde \alpha$ ($\tilde \tau_n \sim 1/n^{\tilde \alpha}$) is in fact larger than $\tilde \alpha =3/2$ for most $Pe$ (cf. mode-number dependence of relaxations in Fig.~S.3), which leads to an exponent $\alpha'$ somewhat larger than $\alpha'=1/3$, consistent with the results of Fig.~\ref{fig:seg_msd}, where we find $\alpha' \approx 2/5$. Figure~\ref{fig:seg_msd}(b) emphasizes the universality of the internal dynamics with increasing $pL$. The various curves, especially for the MSD in the  center-of-mass reference frame, asymptotically approach a power-law regime with an exponent close to the predicted value. This polymer specific regime is evidently only pronounced for $pL \gtrsim 10^3$. Hence, it is not clearly visible in  Fig.~\ref{fig:msd-HI} of the MSD obtained from simulations.
\item $t/\tilde \tau_1, \gamma_R t \ll   1 $ --- Taylor expansion of the exponential functions in Eq.~\eqref{eq:meansquaredisp_ac} yields
\begin{align} \label{eq:msd_ballistic}
\lla  \lbar{\Delta  \bm r^2_a}(t) \rra  =  \frac{v_0^2 l \gamma_R}{L} \sum_{n=1}^{\infty} \frac{\ttau_n}{1+ \gamma_R \ttau_n } t^2 ,
\end{align}
consistent with the observed ballistic regime in Fig.~\ref{fig:seg_msd}. This  regime and its dependence on activity and polymer properties is in quantitative agreement with the simulation results of Fig.~\ref{fig:msd-HI}.
\item $t \to 0$ --- The MSD is dominated by Eq.~\eqref{eq:meansquaredisp_eq}, and all modes contribute. Setting $\ttau_n = \ttone /n^{3/2}$ and replacing the sum by an integral yields
\begin{align} \label{eq:msd_equil_approx}
\lla  \lbar{\Delta  \bm r^2_0}(t) \rra  =  \frac{2L}{\pi^2 p \mu} \left(\frac{t}{\ttau_1} \right)^{2/3} \int_0^{\infty} dx \frac{1-e^{-x^{3/2}}}{x^2} .
\end{align}
This is the same relation as obtained for a passive system, except that $\mu$ and $\ttone$ depend on activity.  With Eq.~(\ref{eq:omega_limit}), we find the P\'eclet-number dependence $\langle \lbar{\Delta  \bm r^2_0}(t) \rangle \sim Pe^{-2/3} (t/\tau_Z)^{2/3}$ for $Pe \gtrsim 50$ and $pL \gg 1$.
\end{itemize}

\section{Summary and Conclusions} \label{sec:summary}

We have presented analytical, numerical, and computer simulation results for the conformational and dynamical properties of active semiflexible polymers in the presence of hydrodynamic interactions. In the simulations, the overdamped dynamics of a bead-spring polymer composed of ABP monomers is studied, with hydrodynamic interactions captured by the Rotne-Prager-Yamakawa hydrodynamic tensor. For the analytical treatment, the Gaussian semiflexible polymer model is adopted, which takes into account the polymer inextensibility in a mean-field manner by a constraint for the contour length. Here, activity is modeled as a Gaussian colored noise process with an exponential temporal correlation. Hydrodynamic interactions are taken into account by the preaveraged Oseen tensor. The linearity of the equation of motion allows for its analytical solution. In any case, our active forces do not generate a Stokeslet, higher-order active multipole flow fields, which decay spatial as $1/r^{2}$  or faster with distance from the self-propelled active site, are neglected, and only the presumably dominant Stokeslet field resulting from  intramolecular forces are taken into account.

Most remarkably, we find a strong influence of hydrodynamics on the polymer conformational properties. In absence of hydrodynamics, active flexible polymers (ABPOs-HI), with $pL>10$, monotonically swell with increasing activity, $Pe$, whereas semiflexible polymers, with $pL<10$, shrink at moderate $Pe$ (the actual $Pe$-range depends on the polymer length) and swell for higher $Pe$, similar to flexible polymers \cite{eise:16,eise:17}. In contrast, active polymers in the presence of hydrodynamic interactions (ABPOs+HI) always shrink for moderate $Pe$ independent of stiffness, and swell again for high activities, where the asymptotic extension for $Pe \to \infty$ of an ABPO+HI is significantly smaller than that of an ABPO-HI. The observed strong influence of hydrodynamics appears over a P\'eclet-number range well covered by synthetic active colloids, where typically $Pe \lesssim 150$.\cite{bech:16}

Two stochastic processes determine the size and shape of an ABPO+HI---thermal and active fluctuations. Due to the linearity of the analytical equations of motion and the additivity of the noise, the fluctuations lead to additive contributions  to the mean square end-to-end distance, which are, however, coupled by the inextensibility of the polymer. The active fluctuations yield a contribution quadratic in the propulsion velocity (or P\'eclet number), similar to the quadratic dependence of the MSD of an ABP \cite{elge:15,bech:16}, which leads to a swelling of the polymer, however, with a $Pe$ dependence smaller than quadratic due to the increase of the stretching coefficient with increasing $Pe$. The polymer inextensibility implies enhanced fluctuations of the thermal part of  $\langle \bm r^2_e\rangle$ by the active noise---expressed by the factor $\mu$---corresponding to a decreasing persistence length with increasing activity associated with a shrinkage of the polymer size.

Qualitatively, the behavior can be understood as follows.
An increasing activity yields an increasing persistent displacement $l_m/l = v_0/\gamma_R l = Pe/2$ of a monomer before it changes its propulsion direction. Hence, any disparity in the propulsion direction is amplified by an increasing $Pe$ and leads to, in average, divergent monomer trajectories and an increasing intramolecular tension reflected in the increasing stretching coefficient $\mu$.  For an ABPO-HI, the competing shrinkage of the thermal part and swelling of the active contribution leads to an overall swelling, since swelling exceeds shrinkage. In case of an ABPO+HI, the reduced swelling can descriptively be understood by fluid-induced collective motions compared to random motions in absence of HI (cf. Fig.~\ref{fig:sketch}). Mathematically, this is reflected by the shorter relaxation times $\tilde \tau_n$ compared to $\tau_n$. To achieve a swelling of an ABPO+HI comparable to that of an ABPO-HI requires larger P\'eclet numbers. As a consequence, $\langle \bm r_e^2 \rangle$ of an ABPO+HI assumes a minimum at intermediate $Pe$.

As mentioned  several times, we do not take into account swimmer-specific flow fields of individual monomers. Nonetheless, the intramolecular forces create complex flow fields---from single monomers to the full filament. Already for a pair of monomers, forces along their bond vector constitutes a force dipole, aside from a potential Stokeslet. In fact, such a force-dipole field could also exist for a passive polymer, but the stronger forces of active monomers increase the dipole field and its relevance for the polymer dynamics significantly. Hence, on larger length scales, embracing more monomers, the overall flow field is rather complex and a large number of hydrodynamic multipoles contribute. This dynamically emerging multipoles are a particular feature of ABPOs+HI and, in their sum, lead to the observed polymer shrinkage.

The polymer dynamics is determined by two relaxation processes, the orientational relaxation of an active site/monomer, and the polymer relaxation. This leads to distinct time regimes in the polymer mean square displacement. At short times $t/\ttone,\ \gamma_Rt \ll 1$, activity leads to a ballistic regime, with an enhanced dynamics compared to a passive polymer. For $t/\ttone \ll 1 \ll \gamma_R t$, the MSD is dominated by the internal dynamics, and a polymer-characteristic subdiffusive regime appears. Again, activity and hydrodynamics play a decisive role, leading to a power-law dependence with an exponent, $\alpha' \approx 2/5$, smaller than that of a passive hydrodynamic polymer. In the asymptotic limit of long times, the enhanced diffusive dynamics is no longer affected by the fluid motion, but rather becomes identical to that of an ABPO-HI.

Our studies predict a substantial effect of hydrodynamic interactions on the properties of active polymers. The shrinkage, even in the presence of excluded-volume interactions, results in an enhanced packing, which might be important for DNA organization within the cell nucleus \cite{smre:17,gana:14}. The actual mechanism of DNA packing is unresolved so far, however, DNA transcription or other local enzymatic processes, e.g., active-loop extrusion \cite{golo:16}, provide a continuous local energy influx, and, hence, a source of nonthermal active noise.
Moreover, hydrodynamic interactions could be involved in the observed subdiffusive dynamics of  chromosomal  loci \cite{webe:10,jave:13}, which is typically related to a  viscoelastic \cite{vand:15} or a fractal environment \cite{tamm:15}. Further experimental studies are necessary to resolve the relevance of the various possible mechanisms affecting the dynamics, such as hydrodynamics, confinement, and viscoelasticity.

\section*{Acknowledgments}
This research was funded by the European Union's Horizon 2020 research and innovation programme under Grant agreement No. 674979-NANOTRANS. Financial support by the Deutsche Forschungsgemeinschaft (DFG) within the priority program SPP 1726 ``Microswimmers-from Single Particle Motion to Collective Behaviour'' is also gratefully acknowledged. Moreover, the authors gratefully acknowledge the computing time granted through JARA-HPC on the supercomputer JURECA at Forschungszentrum J\"ulich.



\balance

\footnotesize{
\bibliographystyle{rsc} 


\providecommand*{\mcitethebibliography}{\thebibliography}
\csname @ifundefined\endcsname{endmcitethebibliography}
{\let\endmcitethebibliography\endthebibliography}{}

\end{document}